\documentclass[structabstract]{aa}
\usepackage{color}

\usepackage{natbib}
\bibpunct{(}{)}{;}{a}{}{,} %to follow the A&A style
\usepackage{graphicx}
\usepackage{txfonts}
\def\kms{km s$^{-1}$}
\def\kmss{km s$^{-1}$\space}

\def\microns{$\mu$m\space}
\def\arcsec{$^{\prime\prime}$}
\def\arcsecs{$^{\prime\prime}$\space}

\def\deg{$^{\circ}$}
\def\degs{$^{\circ}$\space}

\def\h2{H$_2$}
\def\n2h{N$_2$H$^+$}

\def\cii{[C\,{\sc ii}]\space}
\def\ciis{[C\,{\sc ii}]}
\def\nii{[N\,{\sc ii}]\space}
\def\hi{H\,{\sc i}\space}
\def\his{H\,{\sc i}}
\def\hii{H\,{\sc ii}\space}
\def\13co{$^{13}$CO}
\def\c18o{C$^{18}$O}
\def\co{$^{12}$CO\space}
\def\cos{$^{12}$CO}

\def\c+{C$^+$}

\def\h2{H$_2$}

\begin{document}
\title{Internal structure of spiral arms   traced with \ciis: Unraveling the WIM, \his, and molecular emission lanes}
\titlerunning{Spiral arm structure of ionized and molecular gas traced with \cii}
\authorrunning{Velusamy, Langer,  Goldsmith, Pineda}

 %  \subtitle{I. Overviewing the $\kappa$-mechanism}

   \author{T. Velusamy,
   %\inst{1} ,
                  W. D. Langer,
           P. F. Goldsmith,
           \and
           J. L. Pineda
           %\fnmsep\thanks{{\it Herschel} is an ESA space observatory with science instruments provided by European-led Principal Investigator consortia and with important participation from NASA.}
          }

 % \offprints{T.\,Velusamy \email{Thangasamy.Velusamy@jpl.nasa.gov}}

   \institute{Jet Propulsion Laboratory, California Institute of Technology,
              4800 Oak Grove Drive, Pasadena, CA 91109, USA\\
              \email{Thangasamy.Velusamy@jpl.nasa.gov}
%         \and
%             University of Alexandria, Department of Geography, ...\\
%             \email{c.ptolemy@hipparch.uheaven.space}
%             \thanks{Who do we thank?}
             }

   \date{Received 19 February 2015; accepted 9 April 2015}

\abstract{The spiral arm tangencies are  ideal lines of sight in which  to determine the distribution of interstellar gas components in the spiral arms and  study the influence of spiral density waves on the interarm gas in the Milky Way. \cii emission in the tangencies delineates the warm ionized component and the photon dominated regions and is  thus  an important probe of spiral arm structure and dynamics.}
{We aim to  use  \ciis, \his, and \co spectral line maps  of   the Crux, Norma, and  Perseus  tangencies  to
analyze the internal structure of the spiral arms in different gas layers.}
{We use \cii $l$--$V$ maps along with those for \hi and \co
 to derive the  average spectral line  intensity profiles over the longitudinal range of each tangency.
  Using the V$_{LSR}$ of the emission features, we locate the  \ciis, \his, and \co emissions  along a cross cut of  % the profile of
  the spiral  arm.  %from inner to outer edge.
  We use the \cii velocity profile to identify  the compressed  warm ionized medium (WIM) in the spiral arm.}
 {We present a   large scale ($\sim$ 15\deg) position-velocity map of the Galactic plane in \cii from $l$ = 326\fdg6 to 341\fdg4 observed with \textit{Herschel} HIFI.  In the spectral line profiles at the tangencies \cii has two emission peaks, one  associated with the compressed WIM and the other  the molecular gas PDRs. %and PDRs, respectively.
When represented as a  cut across the inner to outer edge of the spiral arm,  the    \ciis--WIM   peak appears  closest to the inner edge while \co and \cii associated with molecular gas are at the outermost edge.  \hi has broader emission with an intermediate peak located  nearer to that of \cos.   }
{The velocity resolved spectral line data of the spiral arm tangencies   unravel  the internal structure in the arms locating the emission lanes within them.
We interpret the excess \cii near the tangent velocities   as shock compression of the WIM induced by the spiral  density waves and as the  innermost edge of spiral arms. For the  Norma  and  Perseus arms,  we estimate  widths  of $\sim$ 250 pc in \ciis--WIM and $\sim$ 400 pc in \co  and  overall  spiral arm widths of $\sim$ 500 pc in \cii and \co emissions; in \hi the widths are $\sim$ 400 pc  and $\sim$ 620 pc for Perseus and Norma, respectively.
The electron densities in the WIM are $\sim$ 0.5 cm$^{-3}$, about an order of magnitude higher than the average for the disk. The enhanced electron density  in the WIM is a result of compression of the WIM by the spiral density wave potential.}


 \maketitle
%
%%%%%%%%%%%%%%%%%%%%%%%%%%%%%%%%
%% HERE STARTS THE INTRODUCTION
%%%%%%%%%%%%%%%%%%%%%%%%%%%%%%%%

%\vspace{-1cm}
\section{Introduction}
\label{sec:introduction}

The large scale structure of spiral arms in the Milky Way has been a subject of great interest for understanding the dynamics of the Galaxy and for interpreting its properties. However, there have been long standing disagreements about the number of arms and their physical parameters.
While a majority of published papers
favor a four-arm structure others   prefer a two-arm structure with a small pitch angle, allowing nearly two
turns of the arms within the solar circle.  In a series of papers \citet{vallee2013,vallee2014apjs,vallee2014b,vallee2014aj}  attempted a statistical modeling analysis of all assembled recent positional data on the Milky Way's spiral arms and observed tangencies in a   number of tracers such as CO, \his, methanol masers, hot and cold dust, and FIR cooling lines, such as \ciis.     The most recent version of his idealized synthesized Galactic map can be found in  \citet{vallee2014apjs}.  Modeling the Galactic spiral structure is based mostly on the data at the spiral arm tangents as observed by different gas tracers and stars. However each of these spiral arm tracers can occupy a separate lane, or layer, across an arm %within its inner and outer edges
\citep[e.g.][]{vallee2014apjs,vallee2014aj}, resulting in an inconsistency among the   various  models  extracted from observational data.

In this paper we present  new \cii spectral line {\it l-V} maps of the Galactic plane   covering  {\it l=}326\fdg6 to 341\fdg4  and  {\it l}=304\fdg9 to 305\fdg9 as  obtained by {\it Herschel}\footnote{{\it Herschel} is an ESA space observatory with science instruments provided by European-led Principal Investigator consortia and with important participation from NASA.} HIFI  On-The-Fly (OTF)  mapping.
 These maps illuminate the structure of different gas components in the spiral arms. Using the \ciis, \hi and \co maps of   the Crux, Norma and start of the Perseus  tangencies,  we derive the %cross sectional
intensity  profiles  of their emissions across these spiral arms and quantify the relative displacement
 of the compressed warm ionized medium (WIM), atomic, and molecular gas lanes with respect to the inner and outer edges of the arms.   Our results reveal an evolutionary transition from the lowest to the highest density states induced by the spiral arm potential. This compressed WIM component traced by \cii is   distinct  from the ionized gas   in
  \hii regions which traces the spiral arms with   characteristics similar to those of  molecular gas traced by CO \citep[cf.][]{vallee2014apjs,downes1980}.

The spiral tangent regions \citep[c.f.][]{vallee2008,benjamin2009} are  ideal laboratories in which to study the interaction of the interstellar gas and spiral density waves in the Milky Way. The tangents provide a unique viewing geometry with sufficiently long path lengths   in  relatively narrow velocity ranges  to detect  the diffuse  WIM component traced by \cii   emission   and to study its relationship to the neutral  \hi and molecular \co gas components within spiral arms  and the influence of spiral arm density waves on  the interstellar medium (ISM).

COBE FIRAS observed strong \cii and \nii emission along the Galactic spiral arm tangencies and \citet{steiman2010} fit the COBE results with four well-defined logarithmic spiral arms in the gaseous component of the ISM.  However, COBE's 7\degs beam and spectrally unresolved lines preclude obtaining detailed information on the scale and  properties of the gas within  the spiral tangencies, nor can one use the COBE data to separate   the emission  that arises from the Photon Dominated Regions (PDRs) from that in the WIM.  The HIFI  {\bf G}alactic {\bf O}bservations of {\bf T}erahertz {\bf C+} (GOT C+) survey \citep{Langer2010} of the Milky Way also detects the strongest \cii emission near the spiral arm tangential directions \citep[][]{Pineda2013,velusamy2014}. In the velocity resolved HIFI spectra  \citet{velusamy2012} separated the WIM component of the \cii emission, in velocity space, from that in the molecular and neutral gas.    They suggested that excitation by electrons in the WIM,  with a density enhanced by the spiral arm potential,   accounts for a low surface brightness \cii excess observed at the tangent velocities along the Scutum-Crux spiral tangency.   To determine whether a similar spatial and density distribution is a general property of Galactic spiral arms   it is important to observe the velocity structure of the \cii emission  in other spiral arm tangencies and compare it with the corresponding \hi and \co emissions.

%Using the 4-arm fit to the COBE  \cii  data  suggests that the in a cross section of the spiral arm with respect to the \co emission the location of \cii emission is displaced  $\sim$ 300pc towards the inner edge, on the near side to the Galactic center (GC) \citep{steiman2010, vallee2014apjs}.
%Using the 4-arm fit to the spectrally unresolved COBE  \cii  data  \citet{steiman2010} and \citet{vallee2014apjs} suggest that  in a cross section of the spiral arm  the location of \cii emission is displaced  $\sim$ 300 pc with respect to the \co emission towards the inner edge, on the near side to the Galactic center (GC).

 Here we present  a large scale ($\sim$ 15\deg) position-velocity map of the Galactic plane in \cii and   derive the following  characteristics of the spiral arm features: the relative locations of the peak emissions  of the  WIM, \his, and molecular gas lanes, including the PDRs, and %derive
the width of each layer.  In addition, we use the \cii emission to derive the mean electron density in the WIM. These results confirm our earlier conclusion  \citep[][]{velusamy2012} that in the velocity profile of \cii emission at the Scutum tangency,  the WIM and molecular gas components of \cii are distinguished  kinematically (appearing at well separated  velocities around the tangent velocity).

 In the analysis presented here  we use the fact that \cii emission can arise in the three major constituents of the interstellar gas, namely, fully or partially ionized gas (as in the WIM), neutral atomic gas (as in HI clouds) and in \h2 molecular gas (as in CO clouds or PDRs) excited, respectively, by electrons, H atoms, and \h2 molecules.  In the velocity resolved HIFI spectra these components are identified as demonstrated in the  GOT C+ results \citep[c.f.][]{Pineda2013,langer2014_II,velusamy2014}. Furthermore, \cite{velusamy2014} show that in the inner Galaxy $\sim$ 60\% of the \cii emission is tracing the \h2 molecular gas while at least 30\% of the   \cii is tracing the WIM and that emission in  \hi excited by H atoms   is not a major contributor. Here, to study spiral arm structure  we identify  two major \cii components,  one component  stems from the WIM and the other from PDRs.
  We find that in the spiral arm tangencies the \cii spectral line data alone can be used to study the relative locations of the WIM and molecular gas PDR layers.

An outline of our paper is as follows. The observations are discussed in Section~\ref{sec:observations}.  In Section~\ref{sec:results} we construct the spatial-velocity maps,     and compare  the distributions of \cii with  \hi  and \co and their relation to the spiral arms.  In Section~\ref{sec:discussion} we analyze the velocity structure of these gas components at the spiral arm tangencies and use it to infer the relative locations of the   different gas ``lanes''  within the spiral arm. Note that in discussing the internal structure of the spiral arms we refer to the emission layers of the gas tracers as ``lanes'',  analogous to the terminology  used  in external spiral galaxies.  We also derive the average electron density in the WIM using the \cii emission and a radiative transfer model.  We summarize our results in Section~\ref{sec:summary}.

%%%%%%%%%%%%%%%%%%%%%%%%%%%%%%%%
%% HERE STARTS THE OBSERVATION SECTION
%%%%%%%%%%%%%%%%%%%%%%%%%%%%%%%%
%% FIGURE 1
%% FIGURE 1

\begin{figure}[!htb]
%\vspace{-0.5cm}
%\centering
%\hspace{-1.0cm}
\includegraphics[scale=0.45,angle=-90]{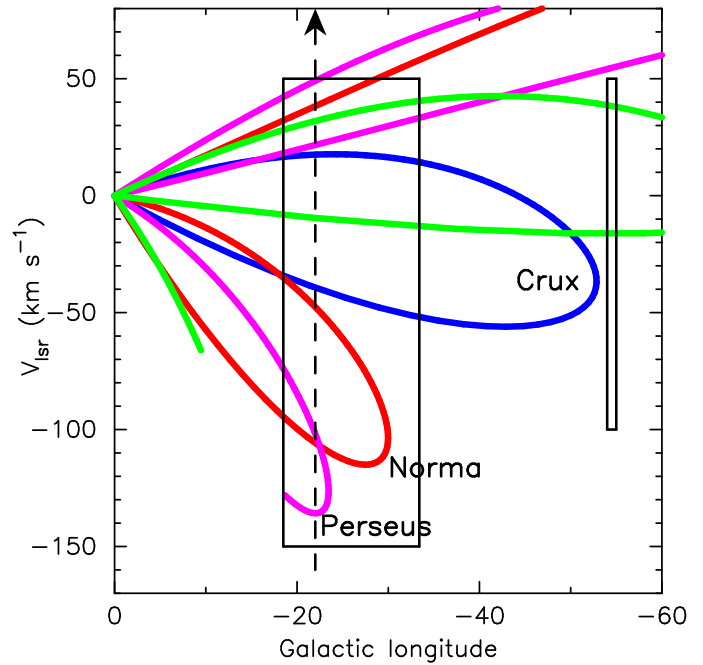}
\caption{Spiral arms in the 4th quadrant are represented in a V$_{LSR}$--longitude ({\it V--l}) plot,  adapted  from \cite{vallee2008}: red: Norma--Cygnus; blue: Scutum--Crux; green: Sagittarius--Carina; magenta: Perseus. The rectangular boxes indicate the  longitude extent in the Galactic plane, at latitude $b$ = 0\deg, of  the HIFI \cii spectral line map data presented here. Note  that the maps cover the  tangencies of the Norma, Crux and start of the Perseus arms. The vertical dashed line represents a line of sight at a given longitude   that intercepts  multiple  spiral arms, thus demonstrating the need for velocity resolved spectral line data to separate them.  }
\label{fig:fig1_vallee}
\vspace{-0.5cm}
\end{figure}

%\vspace{-0.5cm}
\section{Observations}
\label{sec:observations}

The   longitudinal and velocity coverage of the  \cii observations, at latitude $b$ = 0\degs  presented here are summarized in Figure~\ref{fig:fig1_vallee} along with a schematic of the spiral arm velocity--longitude relationship.  All \cii spectral line map observations were made with the high spectral resolution HIFI (de Graauw et al. 2010) instrument onboard {\it Herschel} (Pilbratt et al. 2010). These observations were taken between October 2011 and February 2012.  We used 37 On-The-Fly (OTF) scans for the large scale \cii  map of the Galactic plane (at $b$ = 0\deg) covering a 15\degs range in longitude between 326\fdg6 and 341\fdg4 which include the Norma and Perseus tangencies. The observations   of the fine-structure transition of C$^+$  ($^2$P$_{3/2}$ -- $^2$P$_{1/2}$) at 1900.5369 GHz were made with  the HIFI band 7b   using the wide band spectrometer (WBS). Each OTF scan was    taken  along the Galactic longitude at latitude $b$= 0\degs and was 24 arcmin long.  For the Crux tangency two OTF longitude scan data were used: (i)  one 24 arcmin long centered at {\it l}=305\fdg1 and $b$= 0\degs (note an earlier version of this map was presented in \citet{velusamy2014} and it is included here for completeness in the analysis of the tangencies), and, (ii) a shorter 6 arcmin long scan centered at {\it l}=305\fdg76 and $b$= 0\fdg15.  All HIFI OTF scans were made in the LOAD-CHOP mode using a reference off-source position about 2 degrees  away in latitude (at $b$ = 1\fdg9).  However in our analysis we do not use off-source data (see below).  All  the  24 arcmin long OTF scans are sampled every 40 arcsec  and the shorter scan every 20 arcsec.  The total duration of each OTF scan was typically $\sim$2000 sec which provides only a   short integration time on each spectrum (pixel).  Thus the rms (0.22 K in T$_{mb}$) in the final maps with an  80\arcsecs beam and 2 \kmss wide channels in the OTF spectra is much larger than that in the HIFI spectra observed in the HPOINT mode, for example in the GOT C+ data. The observations for the Crux tangency used  longer integrations.

  We processed the OTF scan map data following the procedure discussed in \citet[][]{velusamy2014}.
The \cii spectral line data  were taken with HIFI Band 7 which utilized Hot Electron Bolometer (HEB) detectors.  These HEBs produced strong electrical standing waves with characteristic periods of $\sim$320 MHz that depend on the signal power.  The HIPE Level 2 \cii spectra  show these residual waves.   We found that applying the {\it fitHifiFringe}\footnote{http://herschel.esac.esa.int/hcss-doc-12.0/index.jsp\#hifi\_um:hifi\-um section 10.3.2}  task to the Level 2 data produced  satisfactory baselines. However, removal of the HEB standing waves has remained a challenge up until the recent release of HIPE-12, which  includes a new tool {\it hebCorrection}\footnote{http://herschel.esac.esa.int/hcss-doc-12.0/index.jsp\#hifi\_um:hifi\-um section 10.4.5} to remove the standing waves in the raw spectral data by  matching the standing wave patterns (appropriate to the power level) in each integration using   a database of spectra at different power levels (see Herschel Science Center (HSC) HIPE-12 release document for details). We used this HSC script to apply {\it hebCorrection} to  create the final pipeline mapping products presented here. Following one of the procedures suggested by
 Dr. David Teyssier  at the HSC, the OTF map data presented here were processed   by re-doing the pipeline  without off source subtraction to produce Level 1 data. The {\it hebCorrection} was then applied to this new Level 1 data. The fact that {\it hebCorrection} subtracts
the matching standing wave patterns from a large database of spectra eliminates the need for off-source subtraction. Thus in our analysis the processed spectral data are free from any off source contamination.  While fitting the HEB waves we also used the feature in the {\it hebCorrection} script to exclude the IF frequencies with  strong \cii emissions.  Finally, the Level 2 data were produced from the HEB corrected Level 1 data.

%% FIGURE 2
%% FIGURE 2

\begin{figure}[!ht]
%\hspace{-0.75cm}
\includegraphics[scale=0.45, angle =-90]{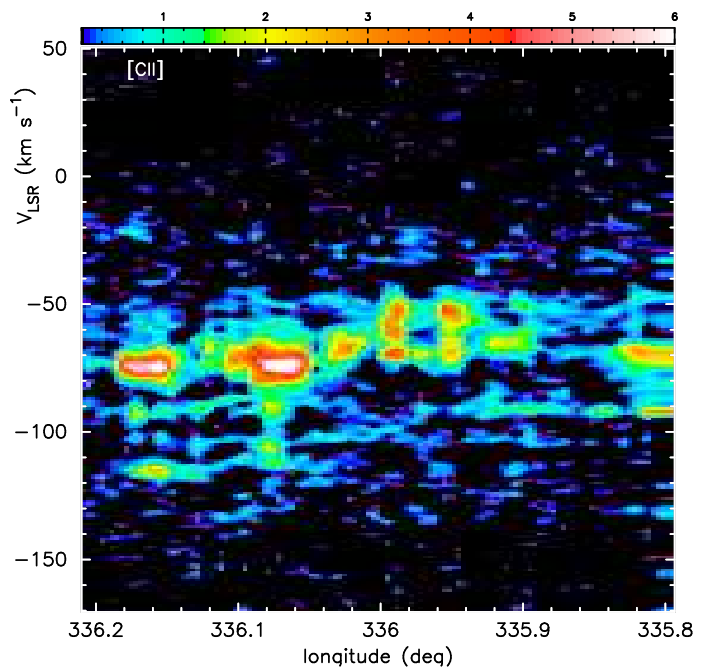}
\caption{Examples of a \cii  OTF longitudinal scan $l$--$V$  map  centered at $l$ = 336\fdg0 and $b$ = 0\fdg0.  The intensities are  in main beam antenna temperature  (T$_{\rm mb}$)   with values  indicated by the color wedge at the top. A square root color stretch is used to bring out the low brightness emission features.    The velocity resolution in all maps is 2 \kmss and the   restored  beam size in longitude is 80\arcsec.}
\label{fig:fig2_otflvmap}
\end{figure}

From Level 2 data the \cii maps were made as ``spectral line cubes'' using the standard mapping scripts in HIPE. Any residual  HEB and optical standing waves in the reprocessed Level 2 data  were minimized further by  applying  {\it fitHifiFringe}   to the ``gridded'' spectral data (we took the additional precaution in {\it fitHifiFringe} of disabling  {\it DoAverage}  in order not to bias the spectral line ``window'').  The H-- and V--polarization data were processed separately and were combined only after applying {\it fitHifiFringe} to the gridded data. This approach minimizes the standing wave residues in the scan maps  by taking into account   the standing wave  differences between H-- and V--polarization.

All  OTF scan map data were reprocessed and analyzed in HIPE 12.1, as described above, to create spectral line data cubes.  We then  used the processed spectral line data cubes to make longitude--velocity ($l$--$V$) maps of   the \cii emission as a function of the longitude range in  each of the 39 OTF observations.     For HIFI observations we used the Wide Band Spectrometer (WBS) with a spectral resolution of 1.1 MHz for all the scan maps. The final $l$--$V$ maps presented here were restored with a velocity resolution of 2 \kms.
 At 1.9 THz the angular resolution of the Herschel telescope is 12\arcsec, but  the \cii OTF observations  used    40\arcsecs sampling. Such fast scanning results in heavily undersampled scans  broadening the effective beam size along the scan direction  \citep[][]{mangum2007}.  Therefore all \cii maps have been   restored with effective beam sizes corresponding to  twice the sampling interval along the scan direction ($\sim$ 80\arcsec). Figure~\ref{fig:fig2_otflvmap} shows an example of a $l$--$V$ map reconstructed using the map data processed in HIPE 12.1 for a single OTF scan map observed at longitude {\it l} = 336\fdg0.

 To compare the distribution of atomic and molecular gas with the ionized gas we use the \co and \hi data  in the southern Galactic plane surveys available in the public archives.    The \cos(1-0)  data are taken from the Three-mm Ultimate Mopra Milky Way Survey\footnote{www.astro.ufl.edu/thrumms}
(ThrUMMS) observed with the 22m Mopra telescope  \citep[][]{Barnes2011}.  The \hi data are taken from the Southern Galactic Plane Survey (SGP) observed with the Australia Telescope Compact Array  \citep[][]{McClure2005}.   Although the Park HIPASS  Galactic plane map of the H 166,167, \& 168$\alpha$ radio recombination lines (RRL)  is now available \citep[][]{alves2015}, its low velocity resolution ($\sim$ 20 \kms) precludes  using it for the analysis presented here.  Furthermore, as the \hii regions are associated with star formation and dense molecular gas \citep[e.g.][]{Anderson2009} the RRL emission from \hii regions is not likely to add more information to  the spiral arm structures than already traced by  \cos.

%% FIGURE 3
%% FIGURE 3

\begin{figure*}[htbp]
%\hspace{-0.75cm}
\includegraphics[scale=0.70, angle=-90]{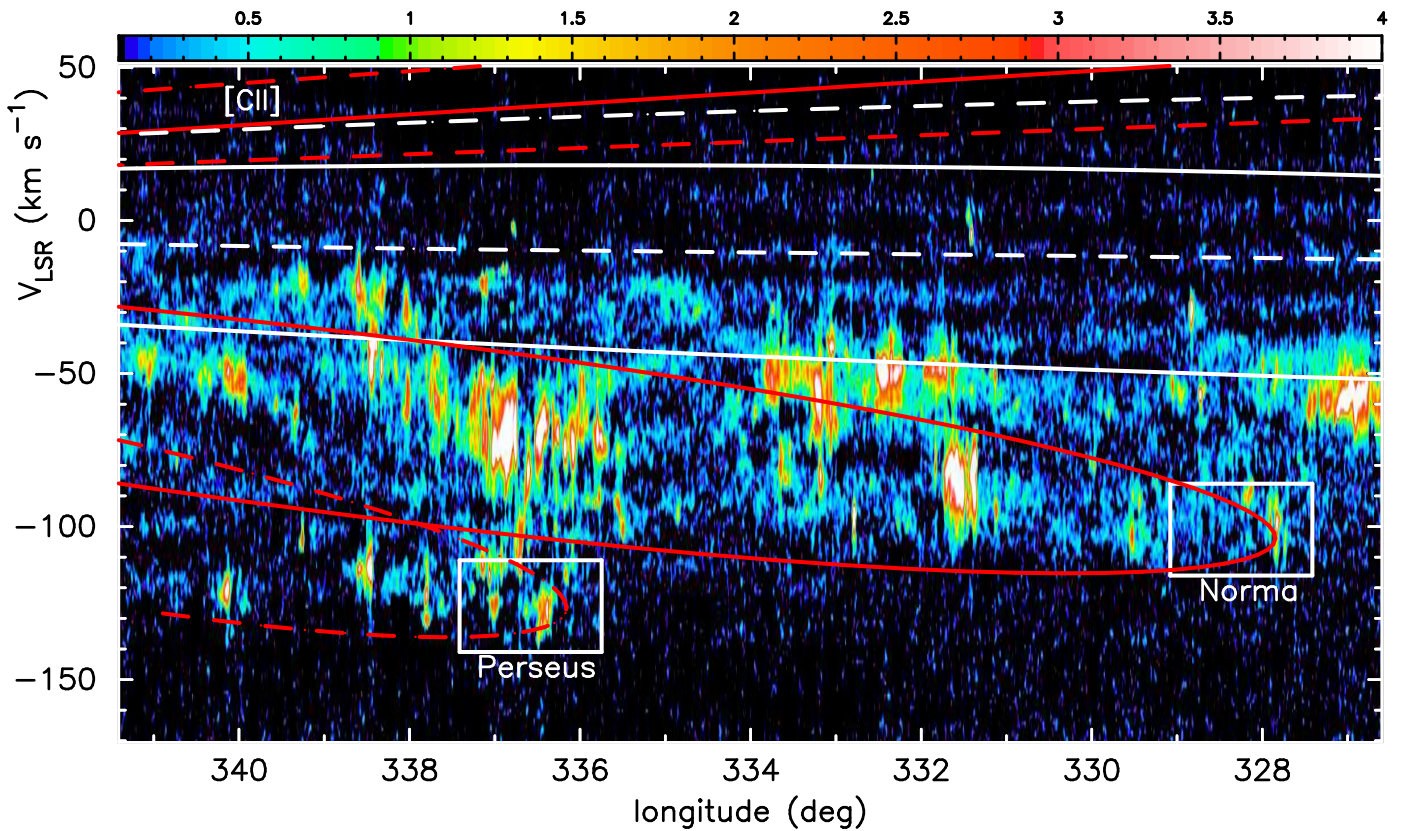}
\caption{The \cii longitude-velocity ({\it l-V})  map covering $\sim$ 15\degs in longitude over the range $l$ = 326\fdg6  to 341\fdg4 at $b$ = 0\fdg0.  The intensities are  in main beam antenna temperature  (T$_{\rm mb}$)   with values  indicated by the color wedge. A square root color stretch is used to bring out the low brightness emission features.    The velocity resolution in all maps is 2 \kmss and the beam size along the longitudinal direction is 80\arcsec. A sketch of the spiral arms in the 4th quadrant,  adopted from \citet{vallee2008}, is overlaid and are indicated in the following colors: red-solid: Norma--Cygnus; white-solid: Scutum--Crux; white-broken: Sagittarius--Carina; red-broken: Perseus.  The rectangular boxes indicate the extent of the spiral tangencies as labeled.}
\label{fig:fig3_[CII]map}
\end{figure*}

%%%%%%%%%%%%%%%%%%%%%%%%%%%%%%%%
%% HERE STARTS THE RESULTS SECTION
%%%%%%%%%%%%%%%%%%%%%%%%%%%%%%%%

%\vspace{-0.5cm}
\section{Results}
\label{sec:results}

In this Section we present the $l$--$V$ emission maps and analyze the structure of the different gas components.  We use a schematic of the expected relationship between  velocity (V$_{LSR}$)  and location with respect to Galactic center, to guide the analysis of the gas lane profile across the arm. We show that the emissions reveal an orderly change in gas components across the arms leading from the least dense WIM to the densest molecular clouds.

\subsection{longitude--velocity maps}

\begin{figure*}[htbp]
%\hspace{-0.75cm}
\includegraphics[scale=0.70, angle=-90]{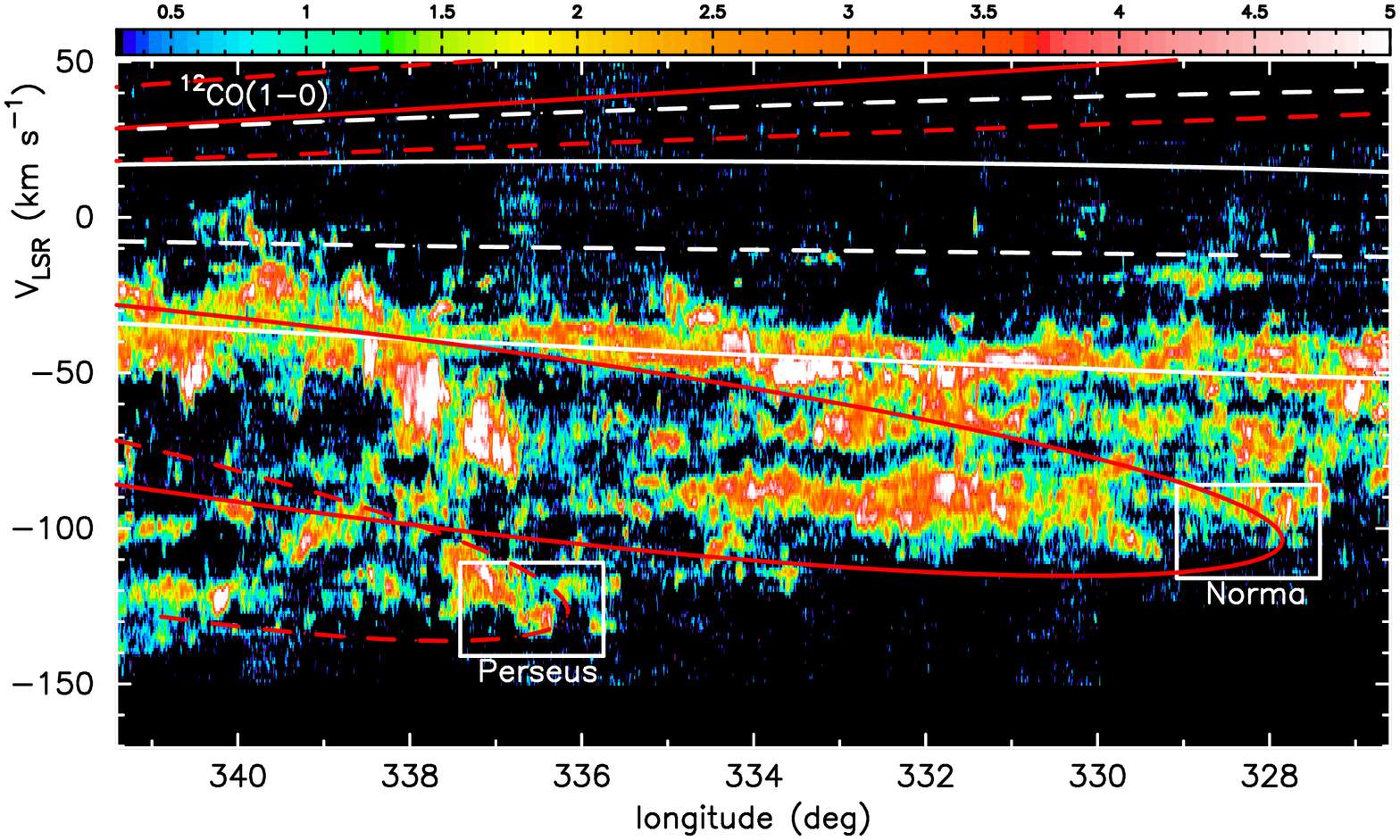}
\caption{The \cos(1-0) longitude--velocity ({\it l-V}) map covering $\sim$ 15\degs in longitude over the range $l$ = 326\fdg6  to 341\fdg4 at $b$ = 0\fdg0.  The \cos(1-0)  data are taken from Mopra ThrUMMS survey \citep{Barnes2011}. See the caption in Figure~\ref{fig:fig3_[CII]map} for the color labels.}
\label{fig:fig 4_COmap}
\end{figure*}

\begin{figure*}[hbp]
%\hspace{-0.75cm}
\includegraphics[scale=0.70, angle=-90]{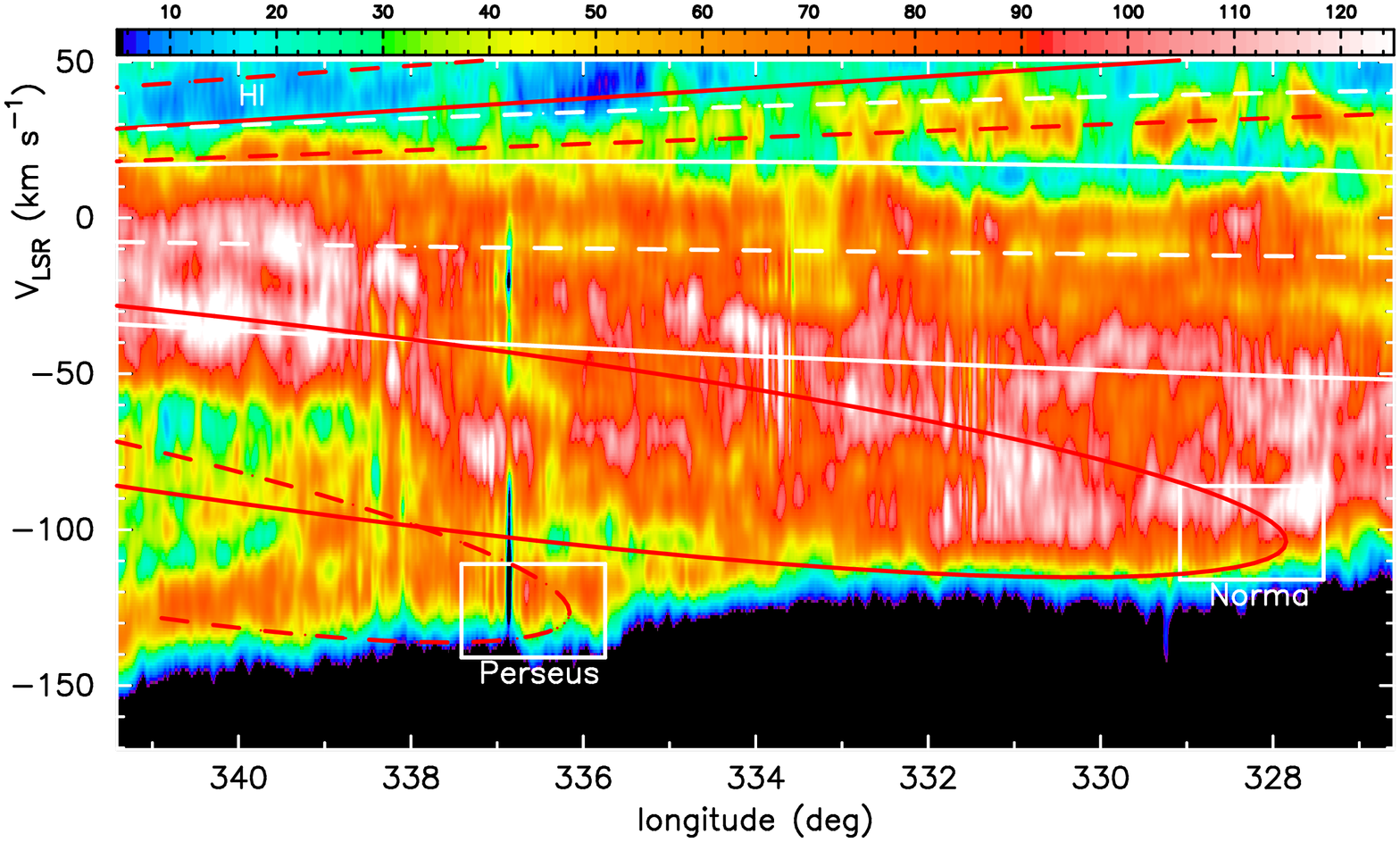}
\caption{The \hi longitude-velocity ({\it l-V}) map covering the $\sim$15\deg range in longitude  $l$ = 326\fdg6  to 341\fdg4 at $b$ = 0\fdg0.  The \hi  data are taken from the SGPS survey \citep{McClure2005}. See the caption in Figure~\ref{fig:fig3_[CII]map} for the color labels.}
\label{fig:fig5_HImap}
\end{figure*}

We ``stitched'' all the individual OTF maps (an example is shown in Figure~\ref{fig:fig2_otflvmap})  within the longitude range 326\fdg6  to 341\fdg4 to create a single longitude velocity map which includes the Perseus and Norma tangencies.  The $l$--$V$ map is shown in Figure~\ref{fig:fig3_[CII]map}.  An  $l$--$V$ representation of the spiral arms is overlaid on the \cii map. Note that strong \cii emissions are seen in the tangencies (denoted by the boxes in Figure~\ref{fig:fig3_[CII]map}) as seen in the GOT C+ survey data \citep[][]{Pineda2013,velusamy2014} and COBE data \citep{steiman2010}.  However,  the rest of the spiral arm trajectories show poor correspondence with the brightness of \cii emission  likely  due to the uncertainties in the model parameters (e.g. pitch angle) used for the spiral arms. We assembled the $l$--$V$ maps that match   the \cii map for the \cos(1-0) (Figure~\ref{fig:fig 4_COmap}) and \hi (Figure~\ref{fig:fig5_HImap}) maps using the Mopra ThrUMMS survey and SGPS data. Note that the intensities for ThrUMMS \cos(1-0) data are uncorrected for main beam efficiency \citep[][]{Ladd2005,Barnes2011}.

%% FIGURE 4
%% FIGURE 4

%% FIGURE 5
%% FIGURE 5

The \cii  maps in Figure~\ref{fig:fig3_[CII]map}  contain,    in addition to information about the tangency, a rich data set  on \cii emission in the diffuse gas,  the molecular gas, and PDRs;  these properties have been analyzed in detail for a sparse galactic sample using the GOT C+ data base \citep[][]{Pineda2013,langer2014_II,velusamy2014}.  However, the continuous longitude coverage in the data   presented here offers a better opportunity  to study the Galactic spiral structure and the internal structure of the arms.  In principle it is possible to derive a 2--D spatial--intensity map of this portion of the Galaxy using the kinematic distances for each velocity feature in the maps. However, such a study is subject to the near-- far--distance ambiguities   inherent in lines of sight inside the solar circle  \citep[see discussion in][]{velusamy2014},   except along the tangencies.  The tangencies offer a unique geometry in which to study kinematics of the interstellar gas without the distance ambiguity.  Furthermore the tangential longitudes provide the longest path length  along the line of sight through  a cross section of the spiral arm thus making it easier to detect weak   \cii emission from the WIM and the molecular gas.  In this paper we limit our analysis only to the \cii emission in tangencies and compare it with those of \hi and \co to understand the structure of the spiral arms in different ISM components.

\subsection{The tangent Emission-velocity profiles}

\begin{figure}[htp]
%\vspace{-0.5cm}
%\centering
\hspace{-1.25cm}
\includegraphics[scale=0.48,angle=0]{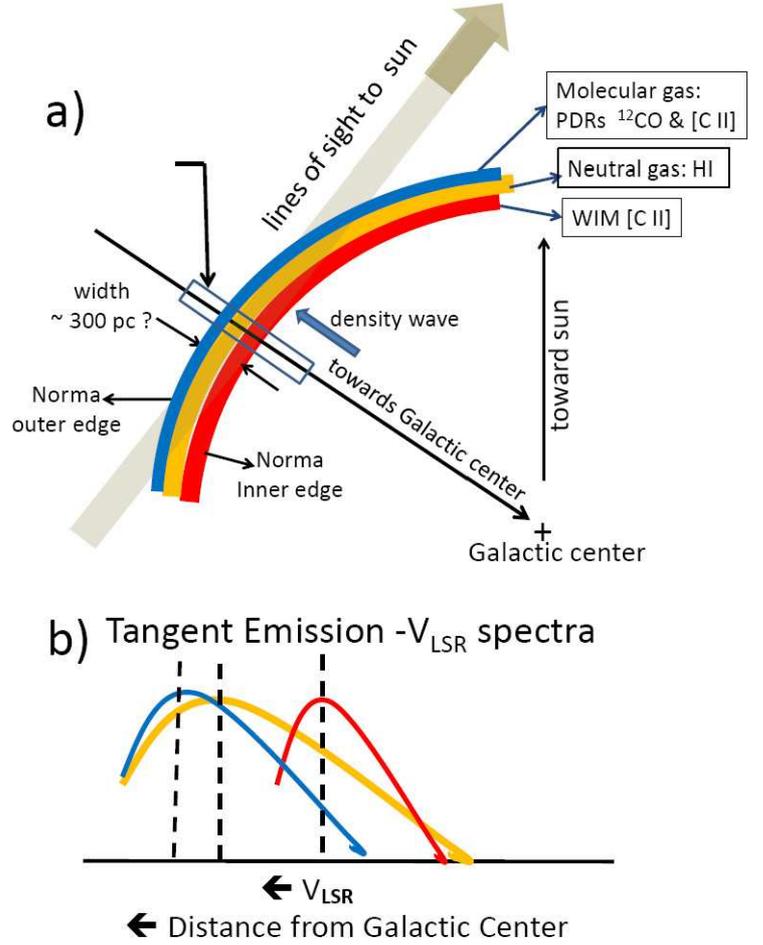}
\caption{ (a) Schematic view of the Norma spiral arm  tangency.   The emissions (distinguished by the color)   tracing the spiral arm  are shown as  a cross cut  of the layers from the inner to outer edges (adapted from Figures 2 \& 3 in \citet[][]{vallee2014apjs}).  (b)  A sketch indicating the  velocity (V$_{LSR}$) structure of the corresponding spectral line intensities near the tangency for each layer. Note that this cartoon  is  intended to be a schematic and is not to scale. }
\label{fig:fig6_l-overview}
\vspace{-0.5cm}
\end{figure}

In Figure~\ref{fig:fig6_l-overview} we show a %toy
schematic  model of the internal structure of a spiral arm  based on the results of \citet{vallee2014apjs} and \citet{velusamy2012}, using the  Norma tangency as an example. The top panel (a) illustrates how the  emission  for the tangency probes the ``lanes'' seen in different tracers.   In Vall\'{e}e's  schematic (see his Figure 5) both \hi and \cii  emissions occur displaced from \co towards the inner edge (on the near side of the  Galactic center). In Vall\'{e}e's sketch the displacement of \cii with respect to \co  is based   the COBE data \citep[][]{steiman2010}. However in the velocity resolved HIFI data for the Scutum tangency \citep[][]{velusamy2012}, the WIM component of \cii emission  occurs near the inner edge while that associated with molecular gas and PDRs is coincident with \co emission. The boxed region in Figure~\ref{fig:fig6_l-overview} represents the area in the  tangency over which the emission spectra are computed. The expected velocity profiles at the tangency, relative   to the  other emission layers, are shown   schematically  in   Figure~\ref{fig:fig6_l-overview}(b) and these     can be compared with the   actual  observed velocity profiles %of the average  emission
for each tracer in the maps in Figures~\ref{fig:fig3_[CII]map} to \ref{fig:fig5_HImap}.    Thus by analyzing    the  velocity profiles of different gas tracers  we can examine the location of the respective emission layers relative to each other  within each spiral arm.

%% FIGURE 6
%% FIGURE 6

 In the $l$--$V$ maps the   trajectory of each spiral arm radial velocity (V$_{LSR}$) as a function of Galactic longitude is a shown as  a  line. In reality, however,  it is expected to be much  broader and complex.    Obviously there is no   single  unique value of the longitude that can be assigned as a  tangency.  Furthermore, the longitudes listed in the literature \citep[e.g.][]{vallee2014apjs} often correspond to data averaged over a wider range of  Galactic latitudes.  But the  $l$--$V$ maps presented here are all in the Galactic plane ($b$ = 0\fdg0) observed with narrow beam sizes (12\arcsec, 33\arcsec, and 150\arcsecs for \ciis, \cos, and \his,  respectively) in latitude. In the  maps  in Figures~\ref{fig:fig3_[CII]map} to \ref{fig:fig5_HImap}  we refer to  a range of longitudes for each spiral tangency as indicated by the box sizes.

\begin{figure}[htp]
\hspace{-0.5cm}
\includegraphics[scale=0.48 ]{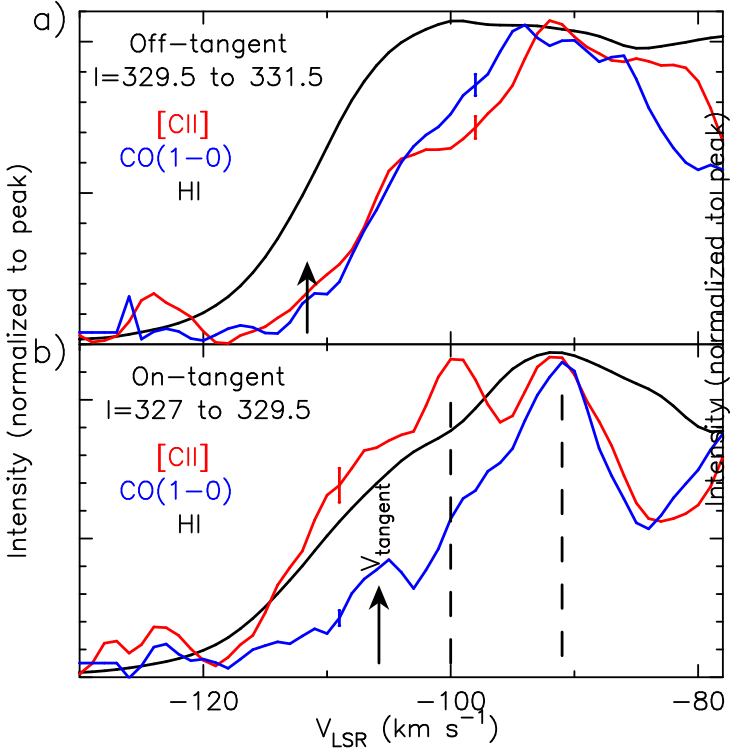}
\caption{Norma tangency spectra.  The \ciis, \his, and \co emission  spectra are plotted against velocity (V$_{LSR}$). Each panel shows the spectra for the longitude ranges indicated.  Note that the intensity scale is normalized to  the peak emission within the velocity range. The corresponding 1--$\sigma$ error bars are indicated on the \cii and \co spectra. The tangent velocity is marked on each panel by a vertical arrow.  Panel (a): Off-tangent. Panel (b): Norma On-tangent.  The dashed lines indicate the V$_{LSR}$ shift between the \cii and \co  peaks.  }
\label{fig:fig7_Norma}
\end{figure}

\begin{figure}[htbp]
\hspace{-0.75cm}
\includegraphics[scale=0.48 ]{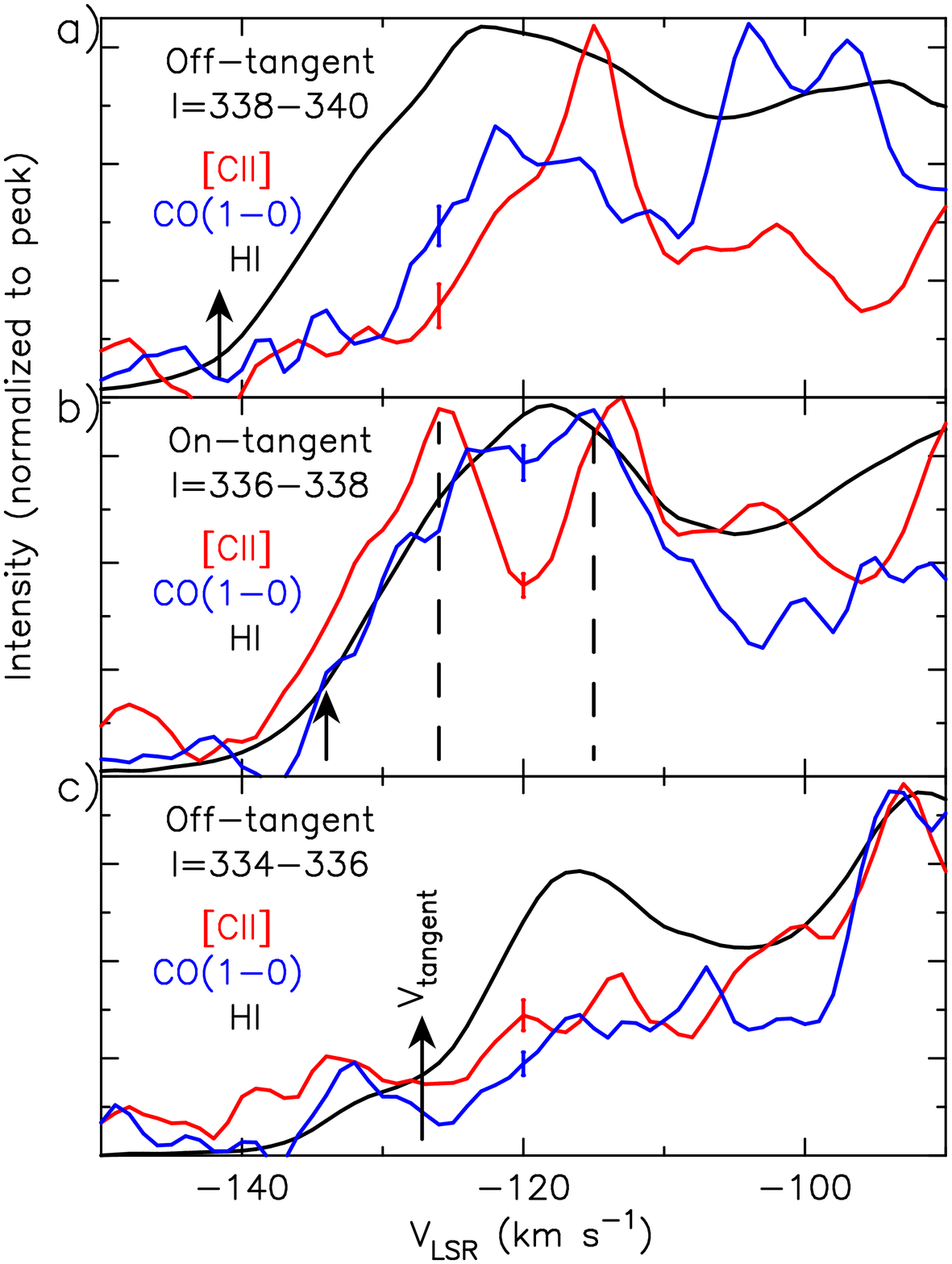}
\caption{Spectra for the location of the start of the Perseus tangency.  Caption same as for Figure~\ref{fig:fig7_Norma}.   Panels (a) \& (c): Off-tangent. Panel (b): Perseus  On-tangent. The dashed lines indicate the shifted V$_{LSR}$ of peak emission of the \cii and \co.     }
\label{fig:fig8_Perseus}
\end{figure}

Using the $l$--$V$ maps in Figures~\ref{fig:fig3_[CII]map} to \ref{fig:fig5_HImap} we average the emissions  within a 2\degs and 2\fdg5   longitude range for the Perseus and Norma tangencies, respectively, and in all observed longitudes for the Crux tangency. The resulting averaged spectra are plotted in Figures~\ref{fig:fig7_Norma}(b) \& \ref{fig:fig8_Perseus}(c).      Note that  the \co spectra shown have new baselines fitted   to   the  map data in Figure~\ref{fig:fig 4_COmap}.  For clarity we limit the velocity range to cover only the tangencies.   Furthermore the intensity scale for each spectrum is adjusted  such that the highest value   corresponds to     its peak brightness  listed in Table~\ref{tab:Table_1}.  The tangent velocity  is indicated on the spectra  in each panel in Figures~\ref{fig:fig7_Norma}, \ref{fig:fig8_Perseus}, \& \ref{fig:fig10_Crux} in order to provide  a reference to  the V$_{LSR}$ velocities.  The tangent velocities are estimated using the mean longitude and assuming  Galactic rotational velocity (220 \kms) at the tangent points \citep[c.f.][]{Levine2008}.  We note  that the tangent velocity varies from panel to panel corresponding to its longitude range.    The \cii  spectra of the tangencies  show remarkably distinct differences compared to spectra of \hi or \cos.  \ciis, unlike \hi or \cos, shows a clear emission peak near the tangent velocity, well separated from the \hi and \co peaks. To bring out  the uniqueness of such differences in the tangencies in Figures 7 and 8  we compare the spectra at tangencies (labeled ``On-tangent'') with those at neighboring longitudes (labeled ``Off-tangent'').  As discussed below, only  the spectra at the tangencies show the excess \cii emission peak  at more negative velocities.    As illustrated in the schematic in Figure~\ref{fig:fig6_l-overview}, %because of the uniqueness of the viewing geometry at the tangencies
it is possible to delineate  individually  the emission layers   or the lanes, bringing out the internal structure of the  spiral arms,  for \cii in both the WIM and the PDRs,  as well as that of molecular gas in \co and atomic gas in \his.     The characteristics of the observed velocity profiles in each tangency are summarized Table~\ref{tab:Table_1}.

%% FIGURE 7
%% FIGURE 7

%% FIGURE 8
%% FIGURE 8

\subsubsection{Norma tangency}

The Norma tangency has been   determined  as 328\degs  for \co \citep[][]{Bronfman2000b}  and \hi \citep[][]{Englmaier1999},   329\degs for 60\microns dust \citep[][]{Bloemen1990}, and 332\degs for 870\microns dust \citep[][]{Beuther2012}.   \citet{garcia2014} assign tangent directions to the Crux (Centaurus), Norma, and 3 kpc expanding arms    of 310\fdg0, 330\fdg0, and 338\fdg0, respectively, by fitting a logarithmic spiral arm model %to the arms
to the  distribution of Giant Molecular clouds (GMCs).  As discussed above, the detection of \cii  from the WIM is  strongest along the tangencies and
thus is a good discriminator of the tangent direction of a spiral arm.   Thus  the   emission profiles shown in Figure~\ref{fig:fig7_Norma}   provide strong evidence   that  the longitude of the tangent direction is well constrained to $l$ $<$ 329\fdg5.  In the On-tangent spectrum the  \co emission shows  a small peak near the tangent velocity.  However, this feature is relatively  weak when compared to the prominent emission seen in \ciis.
\subsubsection{Start of Perseus tangency}

The tangency at the start of the Perseus   arm  has been   determined  as 336\degs  for \co \citep[][]{Bronfman2000b}, 338\degs for the \cii \& \nii FIR lines \citep[][]{steiman2010}, and 338\degs for 870\microns dust \cite[][]{Beuther2012}. \citet[][]{green2011} suggest that part of the Perseus arm could harbor some of the methanol masers found toward the tangent direction of the 3 kpc expanding arm.   According to
the spiral arm model of   \citet[][]{russeil2003}  the starting point of the Perseus arm would be found in the region between the Norma and the 3 kpc expanding arms. Though there are some doubts about this   longitude  being the start of the Perseus arm   \citep[c.f.][]{green2011}, we   adopt  the  longitude range $l$ = 336\deg -- 338\degs as  the start of the Perseus arm  following  the work of \citet[][]{vallee2014apjs}.  The emission profiles  in Figure~\ref{fig:fig8_Perseus} clearly show the detection of a tangent direction in this longitude range, as seen by the strong \ciis--WIM emission compared to the weaker \cii emission in the neighboring longitudes on either side.

\subsubsection{Crux tangency}

\begin{figure}[!htbp]
%\hspace{-0.75cm}
\includegraphics[scale=0.46]{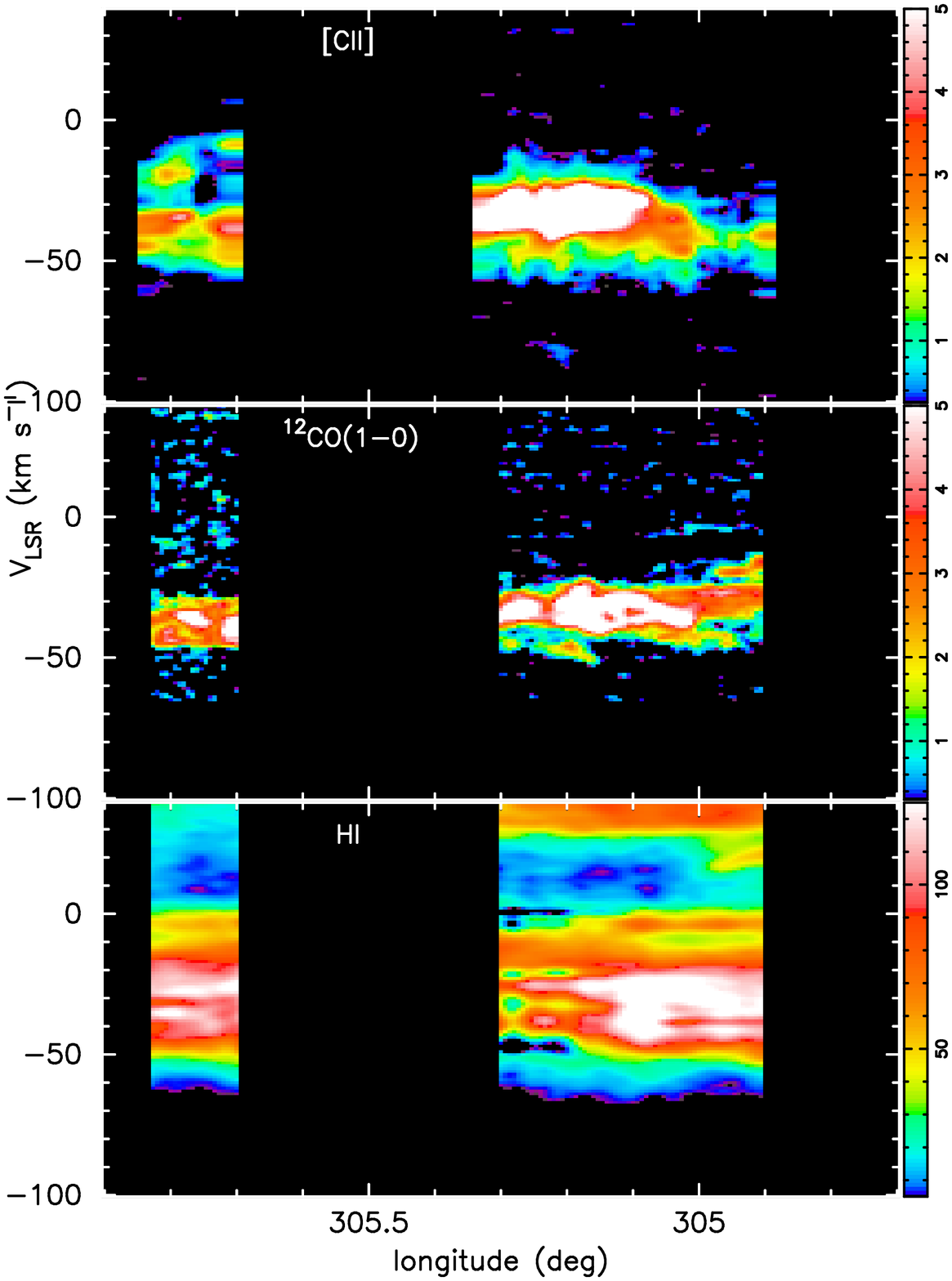}
\caption{The longitude-velocity ({\it l-V}) map covering part of the Crux tangency between longitudes $l$ = 304\fdg9 to 305\fdg9. (a) HIFI \cii maps covered in two OTF longitude scans centered at $l$= 305.1\deg and 305.76\deg at $b$ = 0\fdg0 and 0\fdg15 respectively.  (b) the corresponding \co map from the Mopra ThrUMMS survey. (c) the corresponding  \hi map from SGPS survey. This map region is   off to the right of the  {\it l-V} trajectory of the Crux tangency  as shown in Figure~\ref{fig:fig1_vallee} (see text).  Also see the caption to Figure~\ref{fig:fig3_[CII]map} for details on the display.}
\label{fig:fig9_Crux}
\end{figure}

\citet[][]{vallee2014apjs} lists the Crux\footnote{Also referred to as Centaurus as it appears in this constellation.}  tangency as between  $l$ = 309\degs and 311\degs in different tracers with CO at $l$ = 309\degs  \citep[][]{Bronfman2000b} and  dust at  $l$ = 311\degs \citep[][]{Drimmel2000,Bloemen1990,Beuther2012}.   However, using the GLIMPSE source counts  \citet{benjamin2005} and \citet{churchwell2009} place the Crux tangency   within  a broader longitude range  306\deg $<$  $l$ $<$  313\deg. Furthermore it has been suggested that the Crux tangency provides an ideal testing ground for models of spiral density wave theory \citep[][]{benjamin2008}, as  the $l$ = 302\degs to 313\degs  direction is known to have several distinct anomalies, including large deviations in the \hi velocity field  \citep[][]{McClure2007} and a clear magnetic field reversal  \citep[][]{Brown2007}.   Although we do not have a complete map of this region, it is of sufficient interest that we present a partial map covering $l$ = 304\fdg9 and 305\fdg9, which is close to the Crux tangency.    This  partial  $l$--$V$ map, %covers part of the Crux tangency
 which comes from  another of our \cii {\it Herschel} projects, is shown in Figure~\ref{fig:fig9_Crux} along with the corresponding \co and \hi maps. Note that  the velocity range for the emission in the \co map is much narrower than in the \cii map indicative of a broader diffuse emission component in  \ciis.

%% FIGURE 9
%% FIGURE 9

\begin{figure}[!ht]
%\hspace{-0.5cm}
\includegraphics[scale=0.375, angle = -90 ]{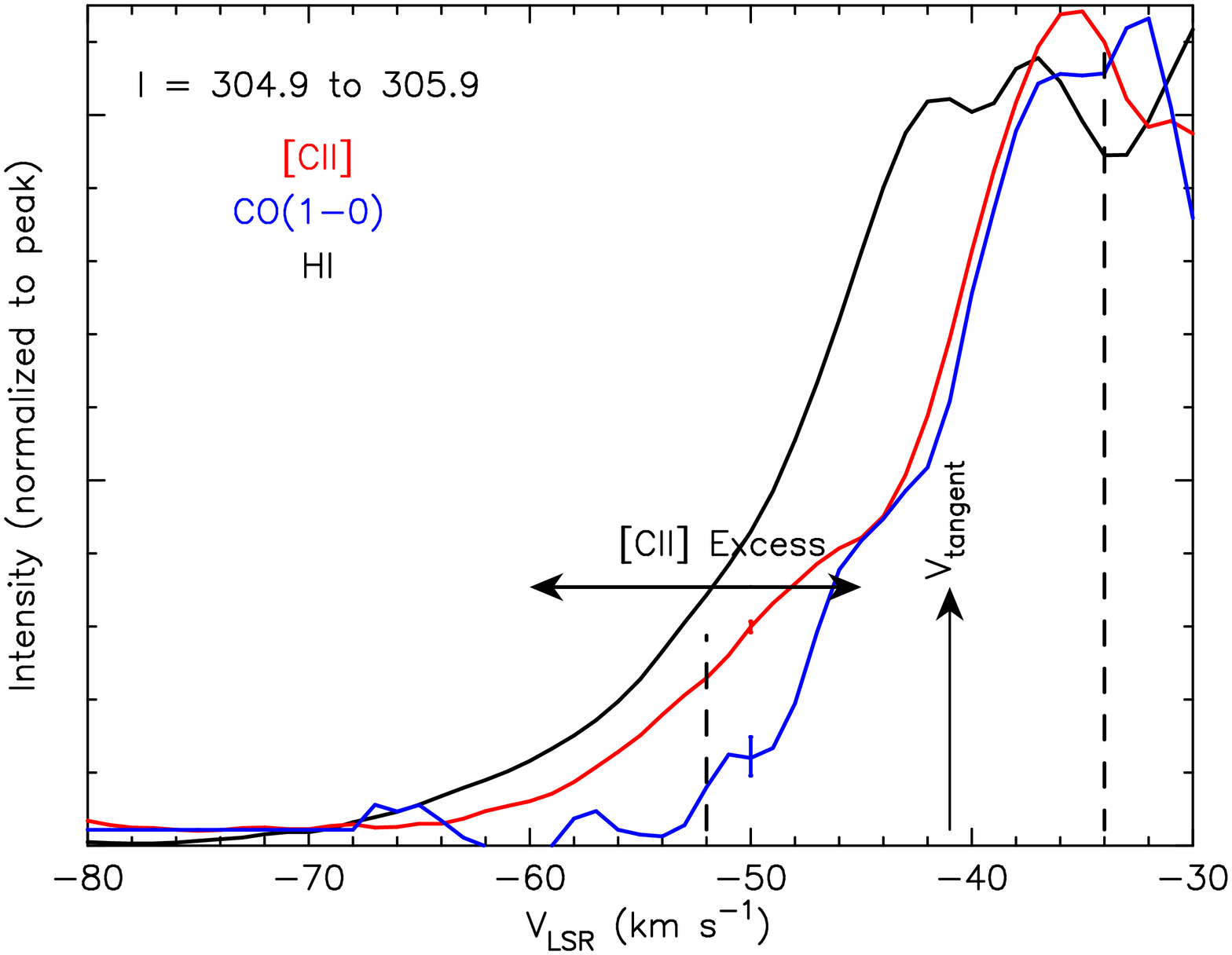}
\caption{ Crux tangency emission spectra showing    \ciis, \his, and \co emission  plotted against velocity (V$_{LSR}$).  The spectra are averaged over  the longitude range indicated.  Note that the intensity scale is normalized to the peak emission within the velocity range. The corresponding 1--$\sigma$ error bars are indicated on the \cii and \co spectra at V$_{LSR}$ = 50 \kms. The tangent velocity is also marked.  The dashed lines indicate the \cii excess beyond the tangent velocity and \co peak in the spectra.  }
\label{fig:fig10_Crux}
\end{figure}

The emission profiles for the Crux tangent region   shown in Figure~\ref{fig:fig10_Crux}  are very similar to those observed for the Scutum tangency \citep[][]{velusamy2014}. The \cii emission shows a clear excess beyond the tangent velocity.  However, unlike the Norma or Perseus tangencies, we do not detect a resolved \cii  emission peak and  the  WIM component appears as   an  enhanced emission shoulder under the \hi   emission profile.  This difference may partly be due to the fact that the longitude range of these maps is well outside the nominal tangent direction and  may also  be due to much stronger emissions seen in both \cii and \co close to the tangent velocity. Nevertheless,  the detection of the \cii excess associated with low \hi and little, or no, \co and its similarity to the results for the Scutum tangency strongly favor its interpretation as the WIM.  This detection of WIM in the longitudes  $l$ = 304\fdg9 to 305\fdg9   indicates  that the Crux tangency is likely to be  much broader ($l$ = 302\degs to 313\deg) in longitude  than the other arms  as was indicated by the   analysis of  star counts in the {\it Spitzer} data \citep[][]{benjamin2005}.

%% FIGURE 10
%% FIGURE 10

%%%%%%%%%%%%%%%%%%%%%%%%%%%%%%%%
%% HERE STARTS THE DISCUSSION SECTION
%%%%%%%%%%%%%%%%%%%%%%%%%%%%%%%%

%\vspace{-0.4cm}
\section{Discussion}
\label{sec:discussion}

The spectra in Figures~\ref{fig:fig7_Norma}, \ref{fig:fig8_Perseus}, and \ref{fig:fig10_Crux} bring out clearly the exceptional characteristics of the \cii emission at the tangencies. These are:
\begin{enumerate}
\item At  the highest velocities beyond the tangent velocity   only \cii shows a peak in emission  while there is little, or no,   \co and \hi is weak, but increasing slowly  with velocity. (The low intensity  \co peak near the tangent velocity in Figure~\ref{fig:fig7_Norma} is relatively less prominent when compared to the dominant \cii emission);
\item  The velocity of the \cii peak  beyond the tangent velocity  corresponds to the radial distance closest to the Galactic center. Therefore  it is near the inner edge of the spiral arm, representing the onset of the spiral arm feature;
\item  The peak emissions for  \hi and \co appear at still higher velocities   than  for the first \cii peak  corresponding to distances away from the inner edge;
\item \cii emission shows two peaks:  one near or beyond the tangent velocity representing the WIM component traced by \cii and  the  second  corresponding to the molecular gas traced by \cii observed in  association with \co arising from the PDRs of the CO clouds; and,
\item The observed velocity profiles are consistent with the schematic   shown in Figure~\ref{fig:fig6_l-overview}, for the internal structure of the spiral arm.
\end{enumerate}

The anomalous excess \cii emission in the velocity profiles for all the spiral arm tangencies represents the direct unambiguous detection of the large scale Galactic diffuse ionized gas (WIM) through its 158\microns [CII] line emission. Our {\it Herschel} HIFI detection of the diffuse ionized gas provides detailed spatial and kinematic information on the nature of this gas component in the spiral arms that has not been possible with prior \cii surveys.  For example, in contrast to the direct detection of the WIM in \cii in our HIFI data,  the deduction of the WIM component in the COBE \cii data \citep[][]{steiman2010} is indirect because it depends on using the  \nii intensities to separate the fraction of \cii intensity from the WIM from other gas components, and is model dependent.

We can be certain that the \cii emissions near the tangent velocity come from the highly ionized WIM and are the result of electron excitation of C$^+$ and not from H atom excitation from diffuse \hi clouds (warm neutral medium or cold neutral medium) for the following reasons, as first noted in our earlier study of  the WIM in the Scutum--Crux arm \citep[][]{velusamy2012,velusamy2014}.
In the spectra in Figures~\ref{fig:fig7_Norma} \& \ref{fig:fig8_Perseus}, the \hi emission at the lowest velocities (near the left around the tangent velocities) is seen along with \cii only for the longitudes of tangent directions which are identified to have \cii in the WIM (Figures~\ref{fig:fig7_Norma}(b) \& \ref{fig:fig8_Perseus}(b)).  However, in all other longitude directions (Figures~\ref{fig:fig7_Norma}(a), \ref{fig:fig8_Perseus}(a) \& \ref{fig:fig8_Perseus}(c)) strong \hi emission is seen with little or no associated \cii emission.  If H atom collisional excitation contributed to any of the \cii emission we  identify  as  coming from the WIM, then  we should have seen \cii associated with the \hi emission for all longitudes in Figures~\ref{fig:fig7_Norma} \& ~\ref{fig:fig8_Perseus}. However, we see none.   It is even more unlikely that \cii excess in the tangents is associated with CO-dark \h2 gas.  In the tangent region the velocity  profiles beyond the tangent velocities the \cii  emission starts appearing at   lower velocities than \cos.  In contrast in the Off-tangent regions in the velocity profiles  both \cii and \co begin to appear simultaneously. Considering the relatively  weak \co emission,  if at all, we expect only a small fraction the \cii excess at the tangent is excited by \h2 molecules. Therefore we assume that \cii excess is dominated by contribution from excitation by electrons.

The detection of WIM emission from our HIFI OTF survey is surprising because the average electron density in the WIM throughout the disk, $\sim$few$\times$10$^{-2}$ cm$^{-3}$, is too low to result in detectable \cii emission at the sensitivity of our HIFI  OTF maps.  Our explanation for the \cii emission detected in our survey is that it originates from denser ionized gas along the inner edge of the spiral arms that has been compressed by the spiral density wave shocks (Figure~\ref{fig:fig6_l-overview}) as previously discussed by \citet{velusamy2012} for the Scutum--Crux arm.
Indeed, as shown below, the electron density is significantly higher in the spiral arm WIM than between the arms.  In what follows we derive the physical characteristics of the spiral arm gas lanes and the electron density in the WIM lane.

\subsection{The internal structure of spiral arms traced by \ciis, \his, and \co}

We can resolve the anatomical structure of the spiral arms, namely   how each gas component is arranged spatially as a function of distance from its edges, using the velocity profiles in the tangencies.   We locate the peak emissions in the layers  traced by \ciis--WIM component, \hi (diffuse atomic clouds), \co (the GMCs), and the \cii in molecular gas (PDRs), as indicated in the cartoon in Figure~\ref{fig:fig6_l-overview},   as a function of radial distance from the Galactic center (GC).  The radial distances are derived from the observed  radial velocities (V$_{LSR}$) assuming a constant Galactic rotation speed of 220 \kmss  for radius $>$ 3 kpc,   at the distance of these spiral arm tangencies. The Galactocentric distances derived from the V$_{LSR}$ are listed in Table~\ref{tab:Table_1}.  The uncertainty in the Galactocentric distances,  which  depends on the assumed Galactic rotational speed   at the tangency, does not affect our results significantly, as we are interested only in the relative displacement between them.    It is easy to characterize the emission profiles   on the  near side to the Galactic Center   near the tangent velocities where the emissions are increasing from zero. On the far side of the Galactic Center the emissions become too complex   to extract the profile parameters due to confusion by emission  from adjacent arms. This   confusion is  especially   prominent  on either side of the peak of the  \cii-WIM component.    Therefore, to derive the width of the emission lanes of each tracer, we use only the cleaner profile on the rising portion of spiral feature on the near side to the Galactic Center.  We compute the Galactocentric radial distances to the V$_{LSR}$ of peaks and the half intensity points of each emission profile.  Using these   radial distances  we  get the distance between the peak and half intensity points in each emission lane. The total width is obtained by simply doubling this value and the results are summarized in Tables~\ref{tab:Table_1} and \ref{tab:Table_2}.

\begin{figure}[!ht]
%\hspace{-0.5cm}
\includegraphics[scale=0.38, angle = -90 ]{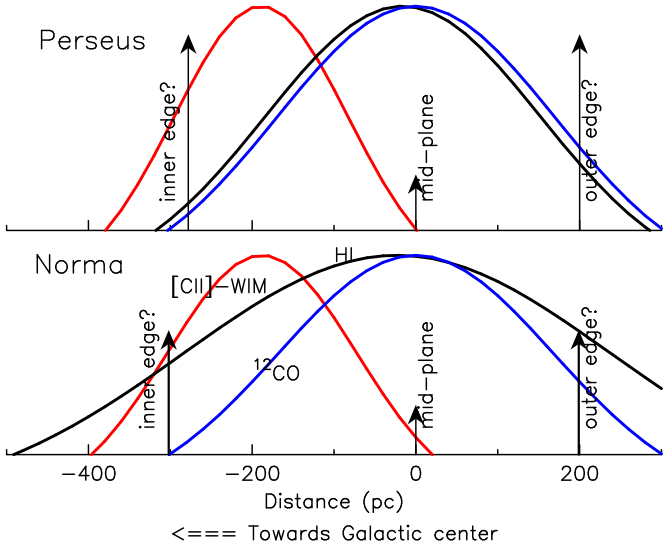}
\caption{Fitted spiral arm structure of the  emission lanes.  Cross cut view of the  structure of the spiral arms Norma and start of Perseus  are shown   including  the relative locations and widths of the emission lanes. The intensity scale is arbitrary. }
\label{fig:fig11_xcut}
\end{figure}

We derive an approximate lower bound on the width for the spiral arm  by assuming \ciis--WIM traces the inner edge and \co the outer edge, and list them in Table~\ref{tab:Table_2}. Note that in the anatomy suggested by  \citet[][]{vallee2014apjs} the hot dust emission traces the inner edge on the near side of the Galactic center with \co tracing the midplane while \cii is in between. Using the parameters in Table~\ref{tab:Table_1} we sketch the cross cut view of the spiral arm in Figure~\ref{fig:fig11_xcut}.   We plot  each lane tracing the \ciis--WIM, \his, and molecular gas (\cos) by an  approximate Gaussian profile.  Note we did not plot the lanes for the Crux arm because the available data do not include the major part of this tangency (see Section~\ref{sec:results}.3).   We list the results in Table 2 as a rough estimate for this tangency because of the poor longitude coverage of the tangency.  Nevertheless, the largest arm width inferred for Crux seems to be consistent with narrower widths for Perseus and Norma.  The Crux spiral tangency is   at  the largest radial distance ($\sim$ 7 kpc) in contrast to  $\sim$ 3.5 -- 4.5 kpc for the other arms studied here.

  The cross-cut  emission profiles  of the internal structures (Figure~\ref{fig:fig11_xcut})   for Norma and the start of the Perseus spiral arms are quite consistent with each other.   The overall sizes  are $\sim$ 480 pc  and $\sim$ 500 pc for the  Perseus and Norma arms, respectively.  To be consistent with  \citet[][]{vallee2014apjs} we present the location of the emission lanes of \ciis--WIM and \hi with respect to  the location of the \co peak emission.   We identify the inner edge of the arm as  the location   of the half power point in the \cii emission profile on the near side of the Galactic Center and on far side from \cos. Similarly we identify the outer edge as  the location   of the half power point in \co emission profile on the far side from the Galactic Center.  We calculate the spiral arm sizes as the distance between inner and outer edges as marked in Figure~\ref{fig:fig11_xcut}. With the exception of the Crux arm (which also has the poor longitude coverage of the tangency in our analysis), the overall sizes  are $\sim$ 480 pc  and $\sim$ 500 pc for the  Perseus and Norma arms, respectively.  The similar width for these arms suggests a width of $\sim$ 500 pc is likely to be characteristic  of all spiral arms in the Galaxy. Our result is close to that obtained by \citet[][]{vallee2014apjs} who estimates a mean width  $\sim$ 600 pc for the  spiral arms using the range of tangent directions observed  using hot dust at the inner edge and \co at the outer edge.  Our analysis and the values  for the widths presented here are likely to be less ambiguous than those of \citet[][]{vallee2014apjs},  as our  data   are based   on   fully sampled longitudinal maps in the galactic plane observed with similar spatial and velocity  resolutions.  Another difference between our data and those used by \citet[][]{vallee2014apjs} is that in his data the tangent directions were observed using maps averaged over a range of latitudes  while ours are only in the plane at $b$ = 0\fdg0.

%% FIGURE 11
%% FIGURE 11

%% TABLE 1
%% TABLE 1

\begin{table}[htbp]
\begin{center}
\caption{Observed spiral arm parameters at the Perseus, Norma and Crux tangencies}
%\vspace{-0.25cm}
\renewcommand*{\arraystretch}{1.4}
\setlength{\tabcolsep}{0.1cm}
%\resizebox{9.0cm}{!} {
\begin{tabular}  {l c c c }
\hline\hline
{\bf Spiral Arm} &	  {\bf Perseus}	&  {\bf Norma}  	&	{\bf Crux}$^1$  	\\
\hline
\multicolumn {4}{l}{Tangency}\\
longitude $l$ = 	&  336\fdg0 - 338\fdg0	&	327\fdg0 - 329\fdg5 	&	304\fdg9 - 305\fdg9  	\\
\hline
\multicolumn {4}{l}  {\bf  Emissions: V$_{peak}$ (\kms) } \\
\ciis--WIM &	-126 	&	-99.5 		 	&	-52 	\\
\cii (molecular)$^2$ &  -113 &   -91.5  &-35	\\
\co   & -117  &-91   &     -32.5\\
\hi   &-118   & -92  &      -38\\
\hline
\multicolumn {4}{l} {\bf Emissions: Peak brightness T$_{mb}$(K) }\\
 \ciis--WIM    & 0.56  & 0.35  & $\sim$ 0.94\\
 \ciis(molecular)    & 0.58  & 0.35  & 5.0\\
 \co$^3$ & 2.03 &2.12 & 4.8 \\
 \hi  &70 &118 & 106 \\
 \hline
\multicolumn {4}{l}  {\bf Galactocentric distance of emission layers (kpc)$^4$} \\
  \ciis--WIM  & 3.45  & 4.57  & 6.59  \\
\co (Mid plane) & 3.60 &4.76 &7.20   \\
\hi   &3.58 &4.74 &7.01 \\
\hline\hline
\label{tab:Table_1}
\end{tabular}
%}
\end{center}
\vspace{-0.7cm}
$^1$This longitude range is offset from the true tangency by $\sim$ 2\deg. Our data are available only for this longitude range. \\
$^2$This identification with \h2 gas is used just to distinguish it from the compressed WIM. However, in addition to excitation by \h2  this component may   include some \cii excited by  atomic H or electrons in the diffuse ionized medium. \\
 $^3$Uncorrected for main beam efficiency.\\
$^4$From the Galactic center as determined from the V$_{LSR}$  and Galactic rotation velocity.\\

%\vspace{-0.5cm}
\end{table}

%% TABLE 2
%% TABLE 2

\begin{table}[htbp]
\begin{center}
\caption{Derived Spiral arm structure at the Perseus,  Norma  and Crux tangencies}
%\vspace{-0.25cm}
\renewcommand*{\arraystretch}{1.4}
\setlength{\tabcolsep}{0.1cm}
%\resizebox{9.0cm}{!} {
\begin{tabular}  {l c c c }
\hline\hline
{\bf Spiral Arm} &	  {\bf Perseus}	&  {\bf Norma}  	&	{\bf Crux}$^1$  	\\
%Tangency	&	Perseus	& Norma & Norma 	&	Crux	\\
\hline
\multicolumn {4}{l} {\bf Relative lane location wrt \co (pc)} \\
 \ciis--WIM peak  & -190 & -190 & -600\\
 \hi peak &-20 & -25 & -180 \\
 \hline
\multicolumn {4}{l} {\bf Emission lane width$^2$ (pc)}\\
 \ciis--WIM    & 250 & 270  &  $\sim$ 220\\
 \co & 400 &400 &  $\sim$ 440  \\
 \hi  &400 &620 &  $\sim$ 640  \\
 Full arm width traced by \cii and \co &480 & 500& $\sim$ 940 \\
\hline
\multicolumn {4}{l}  {\bf WIM parameters at the tangency}\\
Width ${\Delta}$V (\kms)  &   16    &    17        &   12\\
Intensity (K \kms) &  5.1    &       4.11       &     13.74\\
Path length$^3$ (kpc)  & 1.31 & 1.58  & 1.72 \\
$<$n(e)$>$$^4$ (cm$^{-3}$) & 0.52  & 0.42 & 0.74 \\
\hline\hline
\label{tab:Table_2}
\end{tabular}
%}
\end{center}
\vspace{-0.7cm}
$^1$This longitude range is offset from the true tangency and therefore the parameters listed are only indicative of the trends within the arm structure.\\
   $^2$FWHM as estimated using the profile on the rising side (near-side to the Galactic Center)--see text. \\
    $^3$Mean path length estimated using the radial distance from the Galactic Center and the width.\\
    $^4$Assuming that the \cii emission here is due to C$^+$  excitation by electrons and that   excitation  by H atoms is negligible \citep[c.f.][]{velusamy2012,velusamy2014}.
%\vspace{-0.5cm}
\end{table}

As noted above, \citet[][]{vallee2014apjs} constructs a cross section view of the spiral arms showing where each spiral arm tracer occurs, based on a single value assigned for the  longitude  of each  tangency for  the tracers.   This approach provided a useful insight into the different  gas lanes  in a cross-cut of the   profile  of the spiral arms.  However in reality, as seen in the maps in Figures~\ref{fig:fig3_[CII]map} to \ref{fig:fig5_HImap}, the emissions occur over a range of longitudes and it is too complex to assign a single longitude as the tangency.  In other words, all tracers have emission  within a range of longitudes representing their tangency.
Therefore we take a different approach to resolve and delineate the emission layers  within each tangency kinematically by studying the velocity structure in each tracer spectrum. As illustrated in the cartoon of the tangency (Figure~\ref{fig:fig6_l-overview}) the emissions from different layers will show separate velocity V$_{LSR}$  due  to  Galactic rotation.

\subsection{\cii emission in the compressed WIM along the spiral arm}

The spectra in Figures~\ref{fig:fig7_Norma} to \ref{fig:fig10_Crux} bring out clearly the exceptional characteristics of the \cii emission at the tangencies.   Namely,     near  the extreme low  velocities (near the inner edge of the spiral arm)  only \cii   shows an emission peak  (representing the onset of the spiral feature) while there is little,   or no,   \co and  the \hi intensity is still increasing   with velocity. In contrast, both \hi and \co   emission peaks  appear at still higher velocities   away from the inner edge.
This anomalous excess \cii emission in the velocity profiles is observed only for %all
 the spiral arm tangencies, where the path lengths are largest,   and its detection  represents the direct unambiguous   identification  of the large scale Galactic diffuse ionized gas (WIM) by the 158\microns \cii line.   %In the COBE data the   identification  of the WIM component in the \cii emission is indirect   because it involves calculating   the fraction of \cii intensity estimated from the \nii intensities \citep[][]{steiman2010}.
 The results presented here corroborate the   previous  detection of the WIM in the velocity resolved HIFI data for the Scutum tangency \cite[][]{velusamy2012}.   Our {\it Herschel} HIFI detection of the diffuse ionized gas provides detailed spatial and kinematic information on the nature of this gas component in the spiral arms.

 It has been suggested that in the WIM any contribution to the \cii emission from excitation   by H  atoms is small and negligible \citep[][]{velusamy2012,velusamy2014}. For the WNM and WIM conditions (T$_{k}$ =8000 K; \citet{wolfire2003}) the critical density for excitation of \cii by H atoms is $\sim$1300 cm$^{-3}$, and for electrons $\sim$ 45 cm$^{-3}$   \citep[see][]{goldsmith2012}. Using the \hi intensities integrated over the line width of the \ciis--WIM component and the path lengths listed in Table~\ref{tab:Table_2},  we estimate a mean H density $\langle n(H)\rangle$ $\sim$ 0.59 cm$^{-3}$ and 0.31 cm$^{-3}$ in the Perseus and Norma tangencies respectively.  As shown in the case  for the Scutum tangency \citep[][]{velusamy2012} such low \hi densities cannot account for the \cii emission detected at the tangencies. Our interpretation that the \cii emissions near the tangent velocity are the result of C$^+$ excitation by the collisions with electrons in the WIM and not with H atoms, is further corroborated by
  strong evidence seen in the data presented here  as discussed above in Section \ref{sec:discussion}.

Although the \cii emission from the compressed WIM is observed only at,   or near,  the tangent longitudes, it is omnipresent all along the spiral arms in all directions.    But, unlike the tangent direction, it is not easy to detect  this enhancement  in the  spiral arms  layers due to: (i) insufficient path length through the WIM:  it  is a factor of 4 to 5 smaller when viewed in any other direction than along the tangency, therefore   much higher sensitivities are required to detect the weaker emission; (ii) at other longitudes,  due to the viewing geometry, the velocities    are  blended with  other components and it is difficult to disentangle the diffuse  WIM emission from \cii emissions from the molecular gas and PDRs.  However it is possible   to separate the emissions  with  spectral line data  obtained with higher  sensitivity and by   making  a spaxel by spaxel comparison with CO emission \citep[e.g.][]{velusamy2014}. Thus, the WIM component  may add significantly to the total \cii luminosity in galaxies, while being difficult to detect along the average line of sight.

\subsubsection{Electron densities in the  spiral arm WIM}

The electron density of the WIM is an important parameter for understanding the conditions in the ISM, such as the pressure and ionization rate. To estimate the electron densities required to produce the observed \cii emission in the WIM  we follow the approach in \citet{velusamy2012}. At the low densities of the diffuse medium the excitation is sub-thermal and the emission is optically thin, therefore the intensity in an ionized gas is given by \cite[see Section 4 in][]{velusamy2012},

\begin{equation}
\langle n(e) \rangle \sim 0.27T_3^{0.18}(I( \mbox{\ciis})/L_{kpc})^{0.5},
\end{equation}

\noindent which assumes a fully ionized gas, a fractional abundance of C$^+$  with respect to the total hydrogen density, n$_t$, X(C$^+$) = 1.4$\times$10$^{-4}$, and where L$_{kpc}$ is the path length in kpc, T$_3$ is the kinetic temperature in 10$^3$ Kelvin, and

\begin{equation}
 I( \mbox{\ciis})=\int T_A( \mbox{\ciis})dv
 \end{equation}

\noindent  is the intensity in K \kms. The \cii intensities given in Table~\ref{tab:Table_2} are  integrated over the velocity widths for the \ciis--WIM profiles.   The path lengths listed in Table~\ref{tab:Table_2} are derived using the Galactocentric radial distance and the thickness of the \cii emission layer, as discussed above, assuming approximately circular geometry at the tangencies.

Using this approach, assuming a fully ionized gas, fractional abundance $x$(e)=n(e)/n$_t$ = 1,  we calculate $\langle n(e)\rangle$  for all three tangencies and find $\langle n({\rm e})\rangle$ in the range 0.42  to 0.74 cm$^{-3}$  (see Table~\ref{tab:Table_2}) for T$_{k}$ = 8000 K.  For a fully ionized gas this implies a total density $n$(H$^+$) = $n$(e). These values are strictly a lower limit if the gas is partially ionized, $x(e)<$1, but only weakly so. %as $\langle n_t\rangle  \propto x(e)^{-0.5}$.
The densities in the WIM at the leading edge of the spiral arms are an order of magnitude higher than the average density in the disk which is dominated by the interarm gas. Our determination of the WIM density from the \cii emission is several times higher than the LOS averaged densities inferred from pulsar dispersion and H$\alpha$ measurements, $n$(e) $\sim$few$\times$10$^{-2}$ cm$^{-3}$  \citep[][]{Haffner2009} and we argue that our larger mean value is a result of compression by the WIM--spiral arm interaction.

%%%%%%%%%%%%%%%%%%%%%%%%%%%%%%%%
%% THE SUMMARY SECTION
%%%%%%%%%%%%%%%%%%%%%%%%%%%%%%%%

\section{Summary:}
\label{sec:summary}

We present large scale \cii spectral line maps of the Galactic plane from $l$ = 326\fdg6 to 341\fdg4 and  $l$ = 304\fdg9  to 305\fdg9   observed with \textit{Herschel} HIFI using On-The-Fly  scans.  All maps are shown as longitude-velocity ($l$--$V$) maps at latitude, $b$= 0\fdg0, except for $l$ = 305\fdg7  to 305\fdg9 for which  $b$= +0\fdg15.  The  \cii $l$--$V$ maps along with those for \hi and \cos, available from southern Galactic plane surveys \citep[][]{Barnes2011,McClure2005},     are used  to analyze the internal structure of the spiral arms as traced by these gas layers in the the Crux ($l$ = 304\fdg9  -- 305\fdg9),  Norma ($l$ = 327\deg -- 329\fdg5)  and start of Perseus ($l$ = 336\deg -- 338\deg) tangencies.   Our key  results are:
\begin{enumerate}
\item   We derive the internal structure  of the spiral arm features using the velocity resolved emission  profiles of \ciis, \his, and \co  averaged  over each tangency. These yield  the relative locations of the peak emissions  of the  compressed WIM, \his, and molecular gas lanes, including the PDRs, and derive the width of each gas ``lane''.
\item We find that \cii emission has two components. At  the extreme velocities beyond the tangent velocity   only \cii shows a peak in emission  while there is little  \co and \hi is weak. This \cii component traces the compressed  WIM and is displaced by about 9 \kmss in V$_{LSR}$ corresponding to  $\sim$ 200 pc towards the inner edge of the spiral arm with respect to \co emission.  The second \cii component is roughly coincident with \co and traces the PDRs of the molecular gas. The WIM and molecular gas components of \cii are distinguished  kinematically (appearing at well separated  velocities around the tangent velocity).  Thus, we find that in the spiral arm tangencies the \cii spectral line data alone can be used to study the relative locations of the WIM and molecular gas PDR layers.
\item  The peak velocity of the \ciis--WIM  component  lies  beyond the tangent velocity  and corresponds to the radial distance closest to the Galactic center.  Thus it is near the inner edge of the spiral arm, representing the onset of the spiral arm feature. Both \hi and \co peak emissions appear at still higher velocities corresponding to distances away from this inner edge. The   \co profile thus defines the outer edge of the spiral arm. We derive  the width of the spiral arm as the distance between the two extremes of the half power points in the \ciis--WIM and  \co emission profiles. We estimate  the spiral arm widths as $\sim$ 500 pc near the start of the Perseus arm,  and  for the Norma arm.
 \item We interpret the excess \cii near the tangent velocities   as shock compression of the WIM induced by the spiral  density waves and as the  innermost edge of spiral arms. We use the \cii intensities and a radiative transfer model to determine the electron densities WIM traced by \ciis. The electron densities in the compressed  WIM are $\sim$ 0.5 cm$^{-3}$, about an order of magnitude higher than the average for the disk. The enhanced electron density  in the WIM is a result of compression of the WIM by the spiral density wave potential.
 \item Finally, we  suggest  that  the WIM component traced by \cii at the spiral arm tangencies exists all along the spiral arms in all directions, but unlike in the tangent direction it is not easy to detect  %this enhancement due to
  because of insufficient path length of C$^+$ across the arms,  and   confusion due to  velocities blended with  other components. Thus, the WIM component    along the spiral arms  may add significantly to the total \cii luminosity in galaxies, while being difficult to detect along the average line of sight.

\end{enumerate}

In this paper, we demonstrated the utility of  spectrally resolved  {\it Herschel} HIFI OTF scan maps of \cii emission  to unravel the internal structure of spiral arms using the velocity resolved spectral line profiles at the spiral arm tangencies.  Our results provide direct observational   evidence of  the cross section view of the spiral arms in contrast to the synthetic model  by  \citet[][]{vallee2014apjs} using the longitude tangents as traced by different tracers.   Combining \nii  with \cii  yields additional constraints \citep[e.g.][]{langer2015,Yildiz2015}  and future \nii spectral line  maps of the spiral arms  are needed to characterize fully the compressed WIM detected here, and the Galactic arm--interarm interactions.

%%%%%%%%%%%%%%%%%%%%%%%%%%%%%%%%
%% THE ACKNOWLEDGEMENTS
%%%%%%%%%%%%%%%%%%%%%%%%%%%%%%%%

\begin{acknowledgements}
 We thank the staffs of the ESA {\it Herschel} Science Centre and NASA {\it Herschel} Science Center, and the HIFI, Instrument Control Centre (ICC)  for their help with the data reduction routines. In addition, we owe a special thanks to Dr. David Teyssier for clarifications regarding the {\it hebCorrection} tool. This work was performed at the Jet Propulsion Laboratory, California Institute of Technology, under contract with the National Aeronautics and Space Administration.  {\copyright}2015 California Institute of Technology: USA Government sponsorship acknowledged.
%\vspace{-0.5cm}
\end{acknowledgements}

%%%%%%%%%%%%%%%%%%%%%%%%%%%%%%%%
%% THE BIBLIOGRAPHY
%%%%%%%%%%%%%%%%%%%%%%%%%%%%%%%%

\bibliographystyle{aa} %style aa.bst
\bibliography{aa_CII_spiral_refs_v3}

\begin{thebibliography}{38}
\expandafter\ifx\csname natexlab\endcsname\relax\def\natexlab#1{#1}\fi

\bibitem[{{Alves} {et~al.}(2014){Alves}, {Calabretta}, {Davies}, {Dickinson},
  {Staveley-Smith}, {Davis}, {Chen}, \& {Barr}}]{alves2015}
{Alves}, M.~I.~R., {Calabretta}, M., {Davies}, R.~D., {et~al.} 2014, ArXiv
  e-prints

\bibitem[{{Anderson} {et~al.}(2009){Anderson}, {Bania}, {Jackson}, {Clemens},
  {Heyer}, {Simon}, {Shah}, \& {Rathborne}}]{Anderson2009}
{Anderson}, L.~D., {Bania}, T.~M., {Jackson}, J.~M., {et~al.} 2009, \apjs, 181,
  255

\bibitem[{{Barnes} {et~al.}(2011){Barnes}, {Yonekura}, {Fukui}, {Miller},
  {M{\"u}hlegger}, {Agars}, {Miyamoto}, {Furukawa}, {Papadopoulos}, {Jones},
  {Hernandez}, {O'Dougherty}, \& {Tan}}]{Barnes2011}
{Barnes}, P.~J., {Yonekura}, Y., {Fukui}, Y., {et~al.} 2011, \apjs, 196, 12

\bibitem[{{Benjamin}(2008)}]{benjamin2008}
{Benjamin}, R.~A. 2008, in Astronomical Society of the Pacific Conference
  Series, Vol. 387, Massive Star Formation: Observations Confront Theory, ed.
  H.~{Beuther}, H.~{Linz}, \& T.~{Henning}, 375

\bibitem[{{Benjamin}(2009)}]{benjamin2009}
{Benjamin}, R.~A. 2009, in IAU Symposium, Vol. 254, IAU Symposium, ed.
  {J.~Andersen, J.~Bland-Hawthorn, \& B.~Nordstr{\"o}m}, 319--322

\bibitem[{{Benjamin} {et~al.}(2005){Benjamin}, {Churchwell}, {Babler},
  {Indebetouw}, {Meade}, {Whitney}, {Watson}, {Wolfire}, {Wolff}, {Ignace},
  {Bania}, {Bracker}, {Clemens}, {Chomiuk}, {Cohen}, {Dickey}, {Jackson},
  {Kobulnicky}, {Mercer}, {Mathis}, {Stolovy}, \& {Uzpen}}]{benjamin2005}
{Benjamin}, R.~A., {Churchwell}, E., {Babler}, B.~L., {et~al.} 2005, \apjl,
  630, L149

\bibitem[{{Beuther} {et~al.}(2012){Beuther}, {Tackenberg}, {Linz}, {Henning},
  {Schuller}, {Wyrowski}, {Schilke}, {Menten}, {Robitaille}, {Walmsley},
  {Bronfman}, {Motte}, {Nguyen-Luong}, \& {Bontemps}}]{Beuther2012}
{Beuther}, H., {Tackenberg}, J., {Linz}, H., {et~al.} 2012, \apj, 747, 43

\bibitem[{{Bloemen} {et~al.}(1990){Bloemen}, {Deul}, \&
  {Thaddeus}}]{Bloemen1990}
{Bloemen}, J.~B.~G.~M., {Deul}, E.~R., \& {Thaddeus}, P. 1990, \aap, 233, 437

\bibitem[{{Bronfman} {et~al.}(2000){Bronfman}, {May}, \&
  {Luna}}]{Bronfman2000b}
{Bronfman}, L., {May}, J., \& {Luna}, A. 2000, in Astronomical Society of the
  Pacific Conference Series, Vol. 217, Imaging at Radio through Submillimeter
  Wavelengths, ed. J.~G. {Mangum} \& S.~J.~E. {Radford}, 66

\bibitem[{{Brown} {et~al.}(2007){Brown}, {Haverkorn}, {Gaensler}, {Taylor},
  {Bizunok}, {McClure-Griffiths}, {Dickey}, \& {Green}}]{Brown2007}
{Brown}, J.~C., {Haverkorn}, M., {Gaensler}, B.~M., {et~al.} 2007, \apj, 663,
  258

\bibitem[{{Churchwell} {et~al.}(2009){Churchwell}, {Babler}, {Meade},
  {Whitney}, {Benjamin}, {Indebetouw}, {Cyganowski}, {Robitaille}, {Povich},
  {Watson}, \& {Bracker}}]{churchwell2009}
{Churchwell}, E., {Babler}, B.~L., {Meade}, M.~R., {et~al.} 2009, \pasp, 121,
  213

\bibitem[{{Downes} {et~al.}(1980){Downes}, {Wilson}, {Bieging}, \&
  {Wink}}]{downes1980}
{Downes}, D., {Wilson}, T.~L., {Bieging}, J., \& {Wink}, J. 1980, \aaps, 40,
  379

\bibitem[{{Drimmel}(2000)}]{Drimmel2000}
{Drimmel}, R. 2000, \aap, 358, L13

\bibitem[{{Englmaier} \& {Gerhard}(1999)}]{Englmaier1999}
{Englmaier}, P. \& {Gerhard}, O. 1999, \mnras, 304, 512

\bibitem[{{Garc{\'{\i}}a} {et~al.}(2014){Garc{\'{\i}}a}, {Bronfman}, {Nyman},
  {Dame}, \& {Luna}}]{garcia2014}
{Garc{\'{\i}}a}, P., {Bronfman}, L., {Nyman}, L.-{\AA}., {Dame}, T.~M., \&
  {Luna}, A. 2014, \apjs, 212, 2

\bibitem[{{Goldsmith} {et~al.}(2012){Goldsmith}, {Langer}, {Pineda}, \&
  {Velusamy}}]{goldsmith2012}
{Goldsmith}, P.~F., {Langer}, W.~D., {Pineda}, J.~L., \& {Velusamy}, T. 2012,
  \apjs, 203, 13

\bibitem[{{Green} {et~al.}(2011){Green}, {Caswell}, {McClure-Griffiths},
  {Avison}, {Breen}, {Burton}, {Ellingsen}, {Fuller}, {Gray}, {Pestalozzi},
  {Thompson}, \& {Voronkov}}]{green2011}
{Green}, J.~A., {Caswell}, J.~L., {McClure-Griffiths}, N.~M., {et~al.} 2011,
  \apj, 733, 27

\bibitem[{{Haffner} {et~al.}(2009){Haffner}, {Dettmar}, {Beckman}, {Wood},
  {Slavin}, {Giammanco}, {Madsen}, {Zurita}, \& {Reynolds}}]{Haffner2009}
{Haffner}, L.~M., {Dettmar}, R.-J., {Beckman}, J.~E., {et~al.} 2009, Reviews of
  Modern Physics, 81, 969

\bibitem[{{Ladd} {et~al.}(2005){Ladd}, {Purcell}, {Wong}, \&
  {Robertson}}]{Ladd2005}
{Ladd}, N., {Purcell}, C., {Wong}, T., \& {Robertson}, S. 2005, \pasa, 22, 62

\bibitem[{{Langer} {et~al.}(2015){Langer}, {Goldsmith}, {Pineda}, {Velusamy},
  {Requena-Torres}, \& {Wiesemeyer}}]{langer2015}
{Langer}, W.~D., {Goldsmith}, P.~F., {Pineda}, J.~L., {et~al.} 2015, \aap, 576,
  A1

\bibitem[{{Langer} {et~al.}(2010){Langer}, {Velusamy}, {Pineda}, {Goldsmith},
  {Li}, \& {Yorke}}]{Langer2010}
{Langer}, W.~D., {Velusamy}, T., {Pineda}, J.~L., {et~al.} 2010, \aap, 521, L17

\bibitem[{{Langer} {et~al.}(2014){Langer}, {Velusamy}, {Pineda}, {Willacy}, \&
  {Goldsmith}}]{langer2014_II}
{Langer}, W.~D., {Velusamy}, T., {Pineda}, J.~L., {Willacy}, K., \&
  {Goldsmith}, P.~F. 2014, \aap, 561, A122

\bibitem[{{Levine} {et~al.}(2008){Levine}, {Heiles}, \& {Blitz}}]{Levine2008}
{Levine}, E.~S., {Heiles}, C., \& {Blitz}, L. 2008, \apj, 679, 1288

\bibitem[{{Mangum} {et~al.}(2007){Mangum}, {Emerson}, \&
  {Greisen}}]{mangum2007}
{Mangum}, J.~G., {Emerson}, D.~T., \& {Greisen}, E.~W. 2007, \aap, 474, 679

\bibitem[{{McClure-Griffiths} {et~al.}(2005){McClure-Griffiths}, {Dickey},
  {Gaensler}, {Green}, {Haverkorn}, \& {Strasser}}]{McClure2005}
{McClure-Griffiths}, N., {Dickey}, J., {Gaensler}, B., {et~al.} 2005, \apjs,
  158, 178

\bibitem[{{McClure-Griffiths} \& {Dickey}(2007)}]{McClure2007}
{McClure-Griffiths}, N.~M. \& {Dickey}, J.~M. 2007, \apj, 671, 427

\bibitem[{{Pineda} {et~al.}(2013){Pineda}, {Langer}, {Velusamy}, \&
  {Goldsmith}}]{Pineda2013}
{Pineda}, J.~L., {Langer}, W.~D., {Velusamy}, T., \& {Goldsmith}, P.~F. 2013,
  \aap, 554, A103

\bibitem[{{Russeil}(2003)}]{russeil2003}
{Russeil}, D. 2003, \aap, 397, 133

\bibitem[{{Steiman-Cameron} {et~al.}(2010){Steiman-Cameron}, {Wolfire}, \&
  {Hollenbach}}]{steiman2010}
{Steiman-Cameron}, T.~Y., {Wolfire}, M., \& {Hollenbach}, D. 2010, \apj, 722,
  1460

\bibitem[{{Vall{\'e}e}(2008)}]{vallee2008}
{Vall{\'e}e}, J.~P. 2008, \aj, 135, 1301

\bibitem[{{Vall{\'e}e}(2013)}]{vallee2013}
{Vall{\'e}e}, J.~P. 2013, International Journal of Astronomy and Astrophysics,
  3, 20

\bibitem[{{Vall{\'e}e}(2014{\natexlab{a}})}]{vallee2014apjs}
{Vall{\'e}e}, J.~P. 2014{\natexlab{a}}, \apjs, 215, 1

\bibitem[{{Vall{\'e}e}(2014{\natexlab{b}})}]{vallee2014b}
{Vall{\'e}e}, J.~P. 2014{\natexlab{b}}, \mnras, 442, 2993

\bibitem[{{Vall{\'e}e}(2014{\natexlab{c}})}]{vallee2014aj}
{Vall{\'e}e}, J.~P. 2014{\natexlab{c}}, \aj, 148, 5

\bibitem[{{Velusamy} \& {Langer}(2014)}]{velusamy2014}
{Velusamy}, T. \& {Langer}, W.~D. 2014, \aap, 572, A45

\bibitem[{{Velusamy} {et~al.}(2012){Velusamy}, {Langer}, {Pineda}, \&
  {Goldsmith}}]{velusamy2012}
{Velusamy}, T., {Langer}, W.~D., {Pineda}, J.~L., \& {Goldsmith}, P.~F. 2012,
  \aap, 541, L10

\bibitem[{{Wolfire} {et~al.}(2003){Wolfire}, {McKee}, {Hollenbach}, \&
  {Tielens}}]{wolfire2003}
{Wolfire}, M.~G., {McKee}, C.~F., {Hollenbach}, D., \& {Tielens}, A.~G.~G.~M.
  2003, \apj, 587, 278

\bibitem[{{Yildiz} {et~al.}(2015){Yildiz}, {Goldsmith}, {Pineda}, \&
  {Langer}}]{Yildiz2015}
{Yildiz}, U., {Goldsmith}, P., {Pineda}, J., \& {Langer}, W. 2015, in American
  Astronomical Society Meeting Abstracts, Vol. 225, American Astronomical
  Society Meeting Abstracts, 451.09

\end{thebibliography}
\end{document}